\documentclass[12pt,a4paper]{article}
\usepackage{ifthen} % for conditional statements
\newboolean{pdflatex}
\setboolean{pdflatex}{false} % False for eps figures
 
\newboolean{articletitles}
\setboolean{articletitles}{true} % False removes titles in references
 
\newboolean{uprightparticles}
\setboolean{uprightparticles}{false} %True for upright particle symbols
 
\newboolean{inbibliography}
\setboolean{inbibliography}{false} %True once you enter the bibliography
 
\usepackage[top=1in, bottom=1.25in, left=1in, right=1in]{geometry}

\usepackage{microtype}
\usepackage{lineno}  % for line numbering during review
\usepackage{xspace} % To avoid problems with missing or double spaces after
\usepackage{caption} %these three command get the figure and table captions automatically small

\usepackage{graphicx}  % to include figures (can also use other packages)
\usepackage{color}
\usepackage{colortbl}
\graphicspath{{./figs/}} % Make Latex search fig subdir for figures

\usepackage{amsmath} % Adds a large collection of math symbols
\usepackage{amssymb}
\usepackage{amsfonts}
\usepackage{upgreek} % Adds in support for greek letters in roman typeset

\newcommand*\patchAmsMathEnvironmentForLineno[1]{%
\expandafter\let\csname old#1\expandafter\endcsname\csname #1\endcsname
\expandafter\let\csname oldend#1\expandafter\endcsname\csname
end#1\endcsname
 \renewenvironment{#1}%
   {\linenomath\csname old#1\endcsname}%
   {\csname oldend#1\endcsname\endlinenomath}%
}
\newcommand*\patchBothAmsMathEnvironmentsForLineno[1]{%
  \patchAmsMathEnvironmentForLineno{#1}%
  \patchAmsMathEnvironmentForLineno{#1*}%
}
\AtBeginDocument{%
\patchBothAmsMathEnvironmentsForLineno{equation}%
\patchBothAmsMathEnvironmentsForLineno{align}%
\patchBothAmsMathEnvironmentsForLineno{flalign}%
\patchBothAmsMathEnvironmentsForLineno{alignat}%
\patchBothAmsMathEnvironmentsForLineno{gather}%
\patchBothAmsMathEnvironmentsForLineno{multline}%
\patchBothAmsMathEnvironmentsForLineno{eqnarray}%
}

\usepackage{hyperref} 
\usepackage[all]{hypcap} % Internal hyperlinks to floats.

\usepackage{xspace} 
\usepackage{upgreek}

\def\lhcb {\mbox{LHCb}\xspace}

\def\MagUp {\mbox{\em Mag\kern -0.05em Up}\xspace}

\ifthenelse{\boolean{uprightparticles}}%
{
 
 \def\Pgamma      {\ensuremath{\upgamma}\xspace}

 \def\Ppi         {\ensuremath{\uppi}\xspace}

 \def\PDelta      {\ensuremath{\Delta}\xspace}                 
 \def\PXi      {\ensuremath{\Xi}\xspace}                 
 \def\PLambda      {\ensuremath{\Lambda}\xspace}                 
 \def\PSigma      {\ensuremath{\Sigma}\xspace}                 
 \def\POmega      {\ensuremath{\Omega}\xspace}                 
 \def\PUpsilon      {\ensuremath{\Upsilon}\xspace}                 
 
 \def\PB      {\ensuremath{\mathrm{B}}\xspace}                 
                  
 \def\PD      {\ensuremath{\mathrm{D}}\xspace}

 \def\PK      {\ensuremath{\mathrm{K}}\xspace}

 \def\Pb      {\ensuremath{\mathrm{b}}\xspace}                 
 \def\Pc      {\ensuremath{\mathrm{c}}\xspace}

 \def\Pi      {\ensuremath{\mathrm{i}}\xspace}

 \def\Ps      {\ensuremath{\mathrm{s}}\xspace}

}
{
 
 \def\Pgamma      {\ensuremath{\gamma}\xspace}

 \def\Ppi         {\ensuremath{\pi}\xspace}

 \mathchardef\PDelta="7101
 \mathchardef\PXi="7104
 \mathchardef\PLambda="7103
 \mathchardef\PSigma="7106
 \mathchardef\POmega="710A
 \mathchardef\PUpsilon="7107
                  
 \def\PB      {\ensuremath{B}\xspace}                 
                  
 \def\PD      {\ensuremath{D}\xspace}

 \def\PK      {\ensuremath{K}\xspace}

 \def\Pb      {\ensuremath{b}\xspace}                 
 \def\Pc      {\ensuremath{c}\xspace}

 \def\Pi      {\ensuremath{i}\xspace}

 \def\Ps      {\ensuremath{s}\xspace}

}

\makeatletter
\ifcase \@ptsize \relax% 10pt
  \newcommand{\miniscule}{\@setfontsize\miniscule{4}{5}}% \tiny: 5/6
\or% 11pt
  \newcommand{\miniscule}{\@setfontsize\miniscule{5}{6}}% \tiny: 6/7
\or% 12pt
  \newcommand{\miniscule}{\@setfontsize\miniscule{5}{6}}% \tiny: 6/7
\fi
\makeatother

\DeclareRobustCommand{\optbar}[1]{\shortstack{{\miniscule (\rule[.5ex]{1.25em}{.18mm})}
  \\ [-.7ex] $#1$}}

\def\g      {{\ensuremath{\Pgamma}}\xspace}

\def\squark    {{\ensuremath{\Ps}}\xspace}

\def\cquark    {{\ensuremath{\Pc}}\xspace}

\def\bquark    {{\ensuremath{\Pb}}\xspace}

\def\pion   {{\ensuremath{\Ppi}}\xspace}
\def\piz    {{\ensuremath{\pion^0}}\xspace}

\def\pip    {{\ensuremath{\pion^+}}\xspace}
\def\pim    {{\ensuremath{\pion^-}}\xspace}
\def\pipm   {{\ensuremath{\pion^\pm}}\xspace}
\def\pimp   {{\ensuremath{\pion^\mp}}\xspace}

\def\kaon    {{\ensuremath{\PK}}\xspace}
  \def\Kbar    {{\kern 0.2em\overline{\kern -0.2em \PK}{}}\xspace}

\def\KorKbar    {\kern 0.18em\optbar{\kern -0.18em K}{}\xspace}

\def\Kp      {{\ensuremath{\kaon^+}}\xspace}
\def\Km      {{\ensuremath{\kaon^-}}\xspace}
\def\Kpm     {{\ensuremath{\kaon^\pm}}\xspace}
\def\Kmp     {{\ensuremath{\kaon^\mp}}\xspace}
\def\KS      {{\ensuremath{\kaon^0_{\mathrm{ \scriptscriptstyle S}}}}\xspace}

  \def\Dbar    {{\kern 0.2em\overline{\kern -0.2em \PD}{}}\xspace}
\def\D       {{\ensuremath{\PD}}\xspace}

\def\DorDbar    {\kern 0.18em\optbar{\kern -0.18em D}{}\xspace}
\def\Dz      {{\ensuremath{\D^0}}\xspace}
\def\Dzb     {{\ensuremath{\Dbar{}^0}}\xspace}

\def\Dm      {{\ensuremath{\D^-}}\xspace}

\def\Dstar   {{\ensuremath{\D^*}}\xspace}

\def\Dstarz  {{\ensuremath{\D^{*0}}}\xspace}

\def\Dstarp  {{\ensuremath{\D^{*+}}}\xspace}

\def\B       {{\ensuremath{\PB}}\xspace}
\def\Bbar    {{\ensuremath{\kern 0.18em\overline{\kern -0.18em \PB}{}}}\xspace}

\def\BorBbar    {\kern 0.18em\optbar{\kern -0.18em B}{}\xspace}
\def\Bz      {{\ensuremath{\B^0}}\xspace}
\def\Bzb     {{\ensuremath{\Bbar{}^0}}\xspace}
\def\Bu      {{\ensuremath{\B^+}}\xspace}
\def\Bub     {{\ensuremath{\B^-}}\xspace}
\def\Bp      {{\ensuremath{\Bu}}\xspace}
\def\Bm      {{\ensuremath{\Bub}}\xspace}
\def\Bpm     {{\ensuremath{\B^\pm}}\xspace}

\def\Bs      {{\ensuremath{\B^0_\squark}}\xspace}
\def\Bsb     {{\ensuremath{\Bbar{}^0_\squark}}\xspace}

  \def\Y#1S{\ensuremath{\PUpsilon{(#1S)}}\xspace}% no space before {...}!

\def\Lz          {{\ensuremath{\PLambda}}\xspace}
\def\Lbar        {{\ensuremath{\kern 0.1em\overline{\kern -0.1em\PLambda}}}\xspace}
\def\LorLbar    {\kern 0.18em\optbar{\kern -0.18em \PLambda}{}\xspace}

\def\Lb      {{\ensuremath{\Lz^0_\bquark}}\xspace}

\def\Lc      {{\ensuremath{\Lz^+_\cquark}}\xspace}

\def\to                 {\ensuremath{\rightarrow}\xspace}

\def\CP                {{\ensuremath{C\!P}}\xspace}

\def\AT#1     {\ensuremath{A_{\mathrm{T}}^{#1}}\xspace}           % 2

\def\C#1      {\ensuremath{\mathcal{C}_{#1}}\xspace}                       % 9
\def\Cp#1     {\ensuremath{\mathcal{C}_{#1}^{'}}\xspace}                    % 7
\def\Ceff#1   {\ensuremath{\mathcal{C}_{#1}^{\mathrm{(eff)}}}\xspace}        % 9  
\def\Cpeff#1  {\ensuremath{\mathcal{C}_{#1}^{'\mathrm{(eff)}}}\xspace}       % 7
\def\Ope#1    {\ensuremath{\mathcal{O}_{#1}}\xspace}                       % 2
\def\Opep#1   {\ensuremath{\mathcal{O}_{#1}^{'}}\xspace}                    % 7

\newcommand{\tev}{\ifthenelse{\boolean{inbibliography}}{\ensuremath{~T\kern -0.05em eV}}{\ensuremath{\mathrm{\,Te\kern -0.1em V}}}\xspace}
\newcommand{\gev}{\ensuremath{\mathrm{\,Ge\kern -0.1em V}}\xspace}
\newcommand{\mev}{\ensuremath{\mathrm{\,Me\kern -0.1em V}}\xspace}
\newcommand{\kev}{\ensuremath{\mathrm{\,ke\kern -0.1em V}}\xspace}
\newcommand{\ev}{\ensuremath{\mathrm{\,e\kern -0.1em V}}\xspace}
\newcommand{\gevc}{\ensuremath{{\mathrm{\,Ge\kern -0.1em V\!/}c}}\xspace}
\newcommand{\mevc}{\ensuremath{{\mathrm{\,Me\kern -0.1em V\!/}c}}\xspace}
\newcommand{\gevcc}{\ensuremath{{\mathrm{\,Ge\kern -0.1em V\!/}c^2}}\xspace}
\newcommand{\gevgevcccc}{\ensuremath{{\mathrm{\,Ge\kern -0.1em V^2\!/}c^4}}\xspace}
\newcommand{\mevcc}{\ensuremath{{\mathrm{\,Me\kern -0.1em V\!/}c^2}}\xspace}

\def\mum  {\ensuremath{{\,\upmu\mathrm{m}}}\xspace}

\def\invfb   {\ensuremath{\mbox{\,fb}^{-1}}\xspace}

\newcommand{\stat}{\ensuremath{\mathrm{\,(stat)}}\xspace}
\newcommand{\syst}{\ensuremath{\mathrm{\,(syst)}}\xspace}

\newcommand{\chisqip}{\ensuremath{\chi^2_{\text{IP}}}\xspace}

\def\gsim{{~\raise.15em\hbox{$>$}\kern-.85em
          \lower.35em\hbox{$\sim$}~}\xspace}
\def\lsim{{~\raise.15em\hbox{$<$}\kern-.85em
          \lower.35em\hbox{$\sim$}~}\xspace}

\def\ptot       {\mbox{$p$}\xspace}
\def\pt         {\mbox{$p_{\mathrm{ T}}$}\xspace}

\def\evtgen     {\mbox{\textsc{EvtGen}}\xspace}

\def\geant      {\mbox{\textsc{Geant4}}\xspace}

\def\photos     {\mbox{\textsc{Photos}}\xspace}

\def\tell1  {TELL1\xspace}
\def\ukl1   {UKL1\xspace}

\usepackage{cite} % Allows for ranges in citations
\usepackage{mciteplus}

\usepackage{longtable} % only for template; not usually to be used in PAPERs
\usepackage{booktabs}
\usepackage{graphicx}
\usepackage{placeins}
 \usepackage[export]{adjustbox}
\usepackage{float}
\floatstyle{plaintop}
\restylefloat{table}
 
\begin{document}
 
 %% Uncomment during review phase.
%% Comment before a final submission.
%\linenumbers
 
%%%%%%%%%%%%%%%%%%%%%%%%%
%%%%% Title     %%%%%%%%%
%%%%%%%%%%%%%%%%%%%%%%%%%
\renewcommand{\thefootnote}{\fnsymbol{footnote}}
\setcounter{footnote}{1}
 
% %%%%%%% CHOOSE TITLE PAGE--------
%\onecolumn

\begin{titlepage}
\pagenumbering{roman}

% Header ---------------------------------------------------
\vspace*{-1.5cm}
\centerline{\large EUROPEAN ORGANIZATION FOR NUCLEAR RESEARCH (CERN)}
\vspace*{1.5cm}
\noindent
\begin{tabular*}{\linewidth}{lc@{\extracolsep{\fill}}r@{\extracolsep{0pt}}}
\vspace*{-1.2cm}\mbox{\!\!\!\includegraphics[width=.12\textwidth]{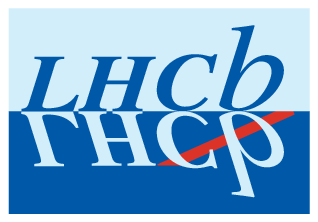}} & &
\\
 & & CERN-EP-2017-195 \\  % ID 
 & & LHCb-PAPER-2017-021 \\  % ID 
 & & December 13, 2017 \\ 
 & & \\
% not in paper \hline
\end{tabular*}

\vspace*{2.0cm}

% Title --------------------------------------------------
{\normalfont\bfseries\boldmath\huge
\begin{center}
Measurement of $C\!P$ observables in $B^\pm \to D^{(*)} K^\pm$ and $B^\pm \to D^{(*)} \pi^\pm$ decays 
\end{center}
}

\vspace*{1.0cm}

% Authors -------------------------------------------------
\begin{center}
The LHCb collaboration\footnote{Authors are listed at the end of this Letter.}
\end{center}

\vspace{\fill}

% Abstract -----------------------------------------------
\begin{abstract}
\noindent
Measurements of $C\!P$ observables in $B^\pm \rightarrow D^{(*)} K^\pm$ and $B^\pm \rightarrow D^{(*)} \pi^\pm$ decays are presented, where $D^{(*)}$ indicates a neutral $\D$ or $\Dstar$ meson that is an admixture of $D^{(*)0}$ and $\bar{D}^{(*)0}$ states. Decays of the $\Dstar$ meson to the $\D\piz$ and $\D\gamma$ final states are partially reconstructed without inclusion of the neutral pion or photon, resulting in distinctive shapes in the \B candidate invariant mass distribution. Decays of the $\D$ meson are fully reconstructed in the $\Kpm \pimp$, $\Kp\Km$ and $\pip\pim$ final states. The analysis uses a sample of charged $B$ mesons produced in $pp$ collisions collected by the LHCb experiment, corresponding to an integrated luminosity of 2.0, 1.0 and 2.0\invfb taken at centre-of-mass energies of \mbox{$\sqrt{s}$ = 7, 8 and 13\tev}, respectively.
The study of $\Bpm \to \Dstar \Kpm$ and $\Bpm \to \Dstar \pipm$ decays using a partial reconstruction method is the first of its kind, while the measurement of $\Bpm \to \D \Kpm$ and $\Bpm \to \D \pipm$ decays is an update of previous LHCb measurements. The $\Bpm \to \D \Kpm$ results are the most precise to date.
\end{abstract}

\vspace*{1.0cm}

\begin{center}
Published in Phys.~Lett.~B777 (2018) 16$-$30
\end{center}

\vspace{\fill}

{\footnotesize 
\centerline{\copyright~CERN on behalf of the \lhcb collaboration, licence \href{http://creativecommons.org/licenses/by/4.0/}{CC-BY-4.0}.}}
\vspace*{2mm}

\end{titlepage}

\newpage
\setcounter{page}{2}
\mbox{~}

\cleardoublepage

%\twocolumn
% %%%%%%%%%%%%% ---------
 
\renewcommand{\thefootnote}{\arabic{footnote}}
\setcounter{footnote}{0}
 
%%%%%%%%%%%%%%%%%%%%%%%%%%%%%%%%
%%%%%  Table of Content   %%%%%%
%%%%%%%%%%%%%%%%%%%%%%%%%%%%%%%%
%%%% Uncomment next 2 lines if desired
%\tableofcontents
%\cleardoublepage

%%%%%%%%%%%%%%%%%%%%%%%%%
%%%%% Main text %%%%%%%%%
%%%%%%%%%%%%%%%%%%%%%%%%%
 
\pagestyle{plain} % restore page numbers for the main text
\setcounter{page}{1}
\pagenumbering{arabic}
 
\section{Introduction}
\label{sec:Introduction}
 
Overconstraining the Unitarity Triangle (UT) derived from the Cabibbo-Kobayashi-Maskawa (CKM) quark-mixing matrix is central to testing the Standard Model (SM) description of \CP violation~\cite{Cabibbo:1963yz,*Kobayashi:1973fv}. The least well known angle of the UT is \mbox{$\gamma \equiv \text{arg}(-V_{ud}V_{ub}^{*}/V_{cd}V_{cb}^{*})$}, which has been determined with a precision of about $7^\circ$ from a combination of measurements~\cite{LHCb-PAPER-2016-032,HFAG} (${cf.}$ $3^\circ$ and $<1^\circ$~on the angles $\alpha$ and $\beta$~\cite{Charles:2004jd,Bona:2006ah}). Among the UT angles, \g is unique in that it does not depend on any top-quark coupling, and can thus be measured in decays that are dominated by tree-level contributions. In such decays, the interpretation of physical observables (rates and \CP asymmetries) in terms of the underlying UT parameters is subject to small theoretical uncertainties~\cite{Brod2014}.  Any disagreement between these measurements of \g and the value inferred from global CKM fits performed without any \g information would invalidate the SM description of \CP violation.

The most powerful method for determining $\gamma$ in decays dominated by tree-level contributions is through the measurement of relative partial widths in $\Bm \to D\Km$ decays, where $D$ represents an admixture of the \Dz and \Dzb states.\footnote{The inclusion of charge-conjugate processes is implied except in any discussion of asymmetries.} 
The amplitude for the $\Bm \to \Dz \Km$ decay, which at the quark level proceeds via a $b \to c\bar{u}s$ transition, is proportional to $V_{cb}$. The corresponding amplitude for the $\Bm \to \Dzb \Km$ decay, which proceeds via a $b\to u\bar{c}s$ transition, is proportional to $V_{ub}$. By studying hadronic \D decays accessible to both \Dz and \Dzb mesons, phase information can be extracted from the interference between these two amplitudes. 
The degree of the resulting \CP violation is governed by the size of $r_B^{DK}$, the ratio of the magnitudes of the $\Bm \to \Dzb \Km$ and $\Bm \to \Dz \Km$ amplitudes. 
The relatively large value of $r_B^{DK} \approx 0.10$~\cite{HFAG} in $\Bm \to D\Km$ decays allows the determination of the relative phase of the two interfering amplitudes. This relative phase has both \CP-violating (\g) and \CP-conserving ($\delta_B^{DK}$) contributions; a measurement of the decay rates for both \Bp and \Bm gives sensitivity to \g. Similar interference effects also occur in $\Bm \to D\pim$ decays, albeit with lower sensitivity to the phases. The reduced sensitivity is the result of additional Cabibbo suppression factors, which decrease the ratio of amplitudes relative to $\Bm \to D\Km$ decays by around a factor of 20.

The $\Bm \to \Dstar \Km$ decay, in which the vector $\Dstar$ meson\footnote{ $\Dstar$ represents an admixture of the $D^{*}(2007)^{0}$ and $\bar{D}^{*}(2007)^{0}$ states.} decays to either the $\D \piz$ or $\D \gamma$ final state, also exhibits \CP-violating effects when hadronic \D decays accessible to both \Dz and \Dzb mesons are studied. In this decay, the exact strong phase difference of $\pi$ between $\Dstar \to \D \piz$ and $\Dstar \to \D \gamma$ decays can be exploited to measure \CP observables for states with opposite \CP eigenvalues~\cite{PhysRevD.70.091503}. The degree of \CP violation observed in $\Bm \to \Dstar \Km$ decays is set by the magnitude of the ratio $r_B^{\Dstar \!K} \approx 0.12$~\cite{HFAG}, and measurement of the phase for both \Bp and \Bm allows \g and $\delta_B^{\Dstar \!K}$ to be disentangled.

The study of $\Bm \to D^{(*)}\Km$ decays for measurements of $\gamma$ was first suggested for \CP eigenstates of the \D decay, for example the \CP-even $\D \to\Kp\Km$ and $\D \to\pip\pim$ decays, labelled here as GLW modes~\cite{Gronau:1990ra,Gronau1991172}. 
In this work, the GLW decays $\D \to \Kp\Km$ and $\D \to \pip\pim$ are considered along with the Cabibbo-favoured $\D \to \Km\pip$ decay, where the latter decay is used for normalisation purposes and to define shape parameters in the fit to data (see Sec.~\ref{sec:Fit}).

The $\Bm \to [h_{1}^+h_{2}^-]_D h^-$ decays, in which $h_{1}^+$, $h_{2}^-$ and $h^-$ can each represent either a charged kaon or pion and the $D$-meson decay products are denoted inside square brackets, have been studied at the \B factories~\cite{Lees:2013nha,PhysRevLett.106.231803} and at LHCb~\cite{LHCb-PAPER-2016-003}. This Letter reports updated and improved results using a sample of charged $B$ mesons from $pp$ collisions collected by the LHCb experiment, corresponding to an integrated luminosity of 2.0, 1.0 and 2.0\invfb taken at centre-of-mass energies of \mbox{$\sqrt{s}$ = 7, 8 and 13\tev}, respectively. The data taken at $\sqrt{s}$ = 13\tev benefits from a higher \Bpm meson production cross-section and a more efficient trigger, so this update of the $\Bm \to [h_{1}^+h_{2}^-]_D h^-$ modes gains approximately a factor of two in signal yield relative to Ref.~\cite{LHCb-PAPER-2016-003}. The $\Bm \to ([h_{1}^+h_{2}^-]_{D} \piz)_{\Dstar}h^-$ and $\Bm \to ([h_{1}^+h_{2}^-]_{D} \gamma)_{\Dstar}h^-$ decays, where the \Dstar-meson decay products are denoted in parentheses, have also been studied by the \B factories~\cite{Aubert:2008ay,Abe:2006hc}, while this work presents the first analysis of these decays at LHCb.

The small $\Dstar - \D$ mass difference and the conservation of angular momentum in \mbox{$\Dstar \to \D\piz$} and \mbox{$\Dstar \to \D\gamma$} decays results in distinctive signatures for the $\Bm \to \Dstar\Km$ signal in the $\D\Km$ invariant mass, allowing yields to be obtained with a partial reconstruction technique.  Since the reconstruction efficiency for low momentum neutral pions and photons is relatively low in LHCb~\cite{LHCb-DP-2014-002}, the partial reconstruction method provides significantly larger yields compared to full reconstruction, but the statistical sensitivity per signal decay is reduced due to the need to distinguish several signal and background components in the same region of $D\Km$ invariant mass.

A total of 19 measurements of \CP observables are reported, eight of which correspond to the fully reconstructed $\Bm \to [h_{1}^+h_{2}^-]_D h^-$ decays while the remaining 11 relate to the partially reconstructed $\Bm \to ([h_{1}^+h_{2}^-]_D \piz/\gamma)_{\Dstar} h^-$ decays. In the latter case, the neutral pion or photon produced in the decay of the \Dstar vector meson is not reconstructed in the final state. A summary of all measured \CP observables is provided in Table~\ref{tab:AllObservables}. In addition, the branching fractions $\mathcal{B}(\Bm \to \Dstarz \pim)$ and $\mathcal{B}(\Dstarz \to \Dz \piz)$, along with the ratio of branching fractions $\frac{\mathcal{B}(\Bm \to \Dstarz \Km)}{\mathcal{B}(\Bm \to \Dz \Km)}$, are reported. 

\begin{table}[!t]
\setlength\extrarowheight{12pt}
\centering
\small
\begin{tabular}{ll}
\toprule
Observable & Definition \\ \hline
$R_{K/\pi}^{K\pi}$ & $\frac{\Gamma(\Bm \to [\Km\pip]_{D}\Km) + \Gamma(\Bp \to [\Kp\pim]_{\D}\Kp)}{\Gamma(\Bm \to [\Km\pip]_{D}\pim) + \Gamma(\Bp \to [\Kp\pim]_{\D}\pip)}$  \\ 
$R^{KK}$ & $\frac{\Gamma(\Bm \to [\Km\Kp]_{D}\Km) + \Gamma(\Bp \to [\Kp\Km]_{\D}\Kp)}{\Gamma(\Bm \to [\Km\Kp]_{D}\pim) + \Gamma(\Bp \to [\Kp\Km]_{\D}\pip)} \times   \frac{1}{R_{K/\pi}^{K\pi}}$  \\
$R^{\pi\pi}$ & $\frac{\Gamma(\Bm \to [\pim\pip]_{D}\Km) + \Gamma(\Bp \to [\pip\pim]_{\D}\Kp)}{\Gamma(\Bm \to [\pim\pip]_{D}\pim) + \Gamma(\Bp \to [\pip\pim]_{\D}\pip)} \times   \frac{1}{R_{K/\pi}^{K\pi}}$  \\
$A_{K}^{K\pi}$ & $\frac{\Gamma(\Bm \to [\Km\pip]_{D}K^{-}) - \Gamma(\Bp \to [\Kp\pim]_{\D}K^{+})}{\Gamma(\Bm \to [\Km\pip]_{D}K^{-}) + \Gamma(\Bp \to [\Kp\pim]_{\D}K^{+})}$ \\
$A_{K}^{KK}$ & $\frac{\Gamma(\Bm \to [\Km\Kp]_{D}K^{-}) - \Gamma(\Bp \to [\Kp\Km]_{\D}K^{+})}{\Gamma(\Bm \to [\Km\Kp]_{D}K^{-}) + \Gamma(\Bp \to [\Kp\Km]_{\D}K^{+})}$  \\
$A_{K}^{\pi\pi}$ & $\frac{\Gamma(\Bm \to [\pim\pip]_{D}K^{-}) - \Gamma(\Bp \to [\pip\pim]_{\D}K^{+})}{\Gamma(\Bm \to [\pim\pip]_{D}K^{-}) + \Gamma(\Bp \to [\pip\pim]_{\D}K^{+})}$  \\
$A_{\pi}^{KK}$ & $\frac{\Gamma(\Bm \to [\Km\Kp]_{D}\pi^{-}) - \Gamma(\Bp \to [\Kp\Km]_{\D}\pi^{+})}{\Gamma(\Bm \to [\Km\Kp]_{D}\pi^{-}) + \Gamma(\Bp \to [\Kp\Km]_{\D}\pi^{+})}$  \\
$A_{\pi}^{\pi\pi}$ & $\frac{\Gamma(\Bm \to [\pim\pip]_{D}\pi^{-}) - \Gamma(\Bp \to [\pip\pim]_{\D}\pi^{+})}{\Gamma(\Bm \to [\pim\pip]_{D}\pi^{-}) + \Gamma(\Bp \to [\pip\pim]_{\D}\pi^{+})}$  \\
$R_{K/\pi}^{K\pi,\piz/\gamma}$ & $\frac{\Gamma(\Bm \to ([\Km\pip]_{D} \piz/\gamma)_{\Dstar} \Km) + \Gamma(\Bp \to ([\Kp\pim]_{D} \piz/\gamma)_{\Dstar} \Kp)}{\Gamma(\Bm \to ([\Km\pip]_{D} \piz/\gamma)_{\Dstar} \pim) + \Gamma(\Bp \to ([\Kp\pim]_{D} \piz/\gamma)_{\Dstar} \pip)}$  \\
$R^{\CP,\piz}$ & $\frac{\Gamma(\Bm \to ([\CP]_{D} \piz)_{\Dstar} \Km) + \Gamma(\Bp \to ([\CP]_{D} \piz)_{\Dstar} \Kp)}{\Gamma(\Bm \to ([\CP]_{D} \piz)_{\Dstar} \pim) + \Gamma(\Bp \to ([\CP]_{D} \piz)_{\Dstar} \pip)} \times \frac{1}{R_{K/\pi}^{K\pi,\piz/\gamma}}$  \\
$R^{\CP,\gamma}$ & $\frac{\Gamma(\Bm \to ([\CP]_{D} \gamma)_{\Dstar} \Km) + \Gamma(\Bp \to ([\CP]_{D} \gamma)_{\Dstar} \Kp)}{\Gamma(\Bm \to ([\CP]_{D} \gamma)_{\Dstar} \pim) + \Gamma(\Bp \to ([\CP]_{D} \gamma)_{\Dstar} \pip)} \times \frac{1}{R_{K/\pi}^{K\pi,\piz/\gamma}}$  \\
$A_{K}^{K\pi,\piz}$ & $\frac{\Gamma(\Bm \to ([\Km\pip]_{D} \piz)_{\Dstar}K^{-}) - \Gamma(\Bp \to ([\Kp\pim]_{\D} \piz)_{\Dstar}K^{+})}{\Gamma(\Bm \to ([\Km\pip]_{D}\piz)_{\Dstar}K^{-}) + \Gamma(\Bp \to ([\Kp\pim]_{\D}\piz)_{\Dstar}K^{+})}$  \\
$A_{\pi}^{K\pi,\piz}$ & $\frac{\Gamma(\Bm \to ([\Km\pip]_{D} \piz)_{\Dstar}\pi^{-}) - \Gamma(\Bp \to ([\Kp\pim]_{\D} \piz)_{\Dstar}\pi^{+})}{\Gamma(\Bm \to ([\Km\pip]_{D}\piz)_{\Dstar}\pi^{-}) + \Gamma(\Bp \to ([\Kp\pim]_{\D}\piz)_{\Dstar}\pi^{+})}$  \\
$A_{K}^{K\pi,\gamma}$ & $\frac{\Gamma(\Bm \to ([\Km\pip]_{D} \gamma)_{\Dstar}K^{-}) - \Gamma(\Bp \to ([\Kp\pim]_{\D} \gamma)_{\Dstar}K^{+})}{\Gamma(\Bm \to ([\Km\pip]_{D}\gamma)_{\Dstar}K^{-}) + \Gamma(\Bp \to ([\Kp\pim]_{\D}\gamma)_{\Dstar}K^{+})}$  \\
$A_{\pi}^{K\pi,\gamma}$ & $\frac{\Gamma(\Bm \to ([\Km\pip]_{D} \gamma)_{\Dstar}\pi^{-}) - \Gamma(\Bp \to ([\Kp\pim]_{\D} \gamma)_{\Dstar}\pi^{+})}{\Gamma(\Bm \to ([\Km\pip]_{D}\gamma)_{\Dstar}\pi^{-}) + \Gamma(\Bp \to ([\Kp\pim]_{\D}\gamma)_{\Dstar}\pi^{+})}$ \\
$A_{K}^{\CP,\piz}$ & $\frac{\Gamma(\Bm \to ([\CP]_{D} \piz)_{\Dstar}K^{-}) - \Gamma(\Bp \to ([\CP]_{\D} \piz)_{\Dstar}K^{+})}{\Gamma(\Bm \to ([\CP]_{D}\piz)_{\Dstar}K^{-}) + \Gamma(\Bp \to ([\CP]_{\D}\piz)_{\Dstar}K^{+})}$  \\
$A_{\pi}^{\CP,\piz}$ & $\frac{\Gamma(\Bm \to ([\CP]_{D} \piz)_{\Dstar}\pi^{-}) - \Gamma(\Bp \to ([\CP]_{\D} \piz)_{\Dstar}\pi^{+})}{\Gamma(\Bm \to ([\CP]_{D}\piz)_{\Dstar}\pi^{-}) + \Gamma(\Bp \to ([\CP]_{\D}\piz)_{\Dstar}\pi^{+})}$  \\
$A_{K}^{\CP,\gamma}$ & $\frac{\Gamma(\Bm \to ([\CP]_{D} \gamma)_{\Dstar}K^{-}) - \Gamma(\Bp \to ([\CP]_{\D} \gamma)_{\Dstar}K^{+})}{\Gamma(\Bm \to ([\CP]_{D}\gamma)_{\Dstar}K^{-}) + \Gamma(\Bp \to ([\CP]_{\D}\gamma)_{\Dstar}K^{+})}$  \\
$A_{\pi}^{\CP,\gamma}$ & $\frac{\Gamma(\Bm \to ([\CP]_{D} \gamma)_{\Dstar}\pi^{-}) - \Gamma(\Bp \to ([\CP]_{\D} \gamma)_{\Dstar}\pi^{+})}{\Gamma(\Bm \to ([\CP]_{D}\gamma)_{\Dstar}\pi^{-}) + \Gamma(\Bp \to ([\CP]_{\D}\gamma)_{\Dstar}\pi^{+})}$  \\
\end{tabular}\captionof{table}{Summary table of the 19 measured \CP observables, defined in terms of $B$ meson decay widths. Where indicated, \CP represents an average of the $\D \to \Kp\Km$ and $\D \to \pip\pim$ modes. The $R$ observables represent partial width ratios and double ratios, where $R_{K/\pi}^{K\pi,\piz/\gamma}$ is an average over the $\Dstar \to \D \piz$ and $\Dstar \to \D\gamma$ modes. The $A$ observables represent \CP asymmetries. \label{tab:AllObservables}}
\end{table}

All of the charge asymmetry measurements are affected by an asymmetry in the \Bpm production cross-section and any charge asymmetry arising from the LHCb detector efficiency, together denoted as $\sigma^\prime$. 
This effective production asymmetry, defined as $A^{\rm eff}_{\Bpm} = \textstyle{\frac{\sigma^\prime(\Bm)-\sigma^\prime(\Bp)}{\sigma^\prime(\Bm)+\sigma^\prime(\Bp)}}$, is measured from the charge asymmetry of the most abundant $\Bm \to [\Km \pip]_{D}\pim$ mode. 
In this mode, the \CP asymmetry is fixed to have the value $A_{\pi}^{K\pi} = (+0.09 \pm 0.05)\%$, which is determined using knowledge of \g and $r_B^{DK}$ from Ref.~\cite{LHCb-PAPER-2016-032}, where $A_{\pi}^{K\pi}$ was not used as an input observable. This uncertainty is smaller than that of previous measurements of the \Bpm production asymmetry measured at $\sqrt{s} = 7$ and 8\tev~\cite{LHCb-PAPER-2016-054,LHCb-PAPER-2016-062}, and reduces the systematic uncertainties of the asymmetries listed in Table~\ref{tab:AllObservables}. The value of $A^{\rm eff}_{\Bpm}$ is applied as a correction to all other charge asymmetries. The remaining detection asymmetries, most notably due to different numbers of \Kp and \Km mesons appearing in each final state, are corrected for using independent calibration samples. These corrections transform the measured charge asymmetries into \CP asymmetries.

\section{Detector and simulation}
\label{sec:Detector}

The \lhcb detector~\cite{Alves:2008zz,LHCb-DP-2014-002} is a single-arm forward
spectrometer covering the \mbox{pseudorapidity} range $2<\eta <5$,
designed for the study of particles containing \bquark or \cquark
quarks. The detector includes a high-precision tracking system
consisting of a silicon-strip vertex detector surrounding the $pp$
interaction region, a large-area silicon-strip detector located
upstream of a dipole magnet with a bending power of about
$4{\mathrm{\,Tm}}$, and three stations of silicon-strip detectors and straw
drift tubes placed downstream of the magnet.
The tracking system provides a measurement of momentum, \ptot, of charged particles with
a relative uncertainty that varies from 0.5\% at low momentum to 1.0\% at 200\gevc.
The minimum distance of a track to a primary vertex (PV), the impact parameter (IP), is measured with a resolution of $(15+29/\pt)\mum$,
where \pt is the component of the momentum transverse to the beam, in\,\gevc.
Different types of charged hadrons are distinguished using information
from two ring-imaging Cherenkov detectors (RICH)~\cite{PAPANESTIS2017,Adinolfi2013}. 
Photons, electrons and hadrons are identified by a calorimeter system consisting of
scintillating-pad and preshower detectors, an electromagnetic
calorimeter and a hadronic calorimeter. Muons are identified by a
system composed of alternating layers of iron and multiwire
proportional chambers.

The trigger consists of a hardware stage, based on information from the calorimeter and muon
systems, followed by a software stage, in which all charged particles
with $\pt>500\,(300)\mev$ are reconstructed for 2011\,(2012) data, and $\pt>70\mev$ for 2015 and 2016 data.
At the hardware trigger stage, events are required to contain a muon with high \pt or a
  hadron, photon or electron with high transverse energy in the calorimeters. For hadrons,
  the transverse energy threshold varied between 3 and 4 GeV between 2011 and 2016.
  The software trigger requires a two-, three- or four-track
  secondary vertex with significant displacement from all primary
  $pp$ interaction vertices. 
  A multivariate algorithm~\cite{BBDT,Likhomanenko:2015aba} is used for
  the identification of secondary vertices consistent with the decay
  of a \bquark hadron.

In the simulation, $pp$ collisions are generated using
\textsc{Pythia8}~\cite{Sjostrand:2007gs}
 with a specific \lhcb
configuration~\cite{LHCb-PROC-2010-056}.  Decays of hadronic particles
are described by \evtgen~\cite{Lange:2001uf}, in which final-state
radiation is generated using \photos~\cite{Golonka:2005pn}. The
interaction of the generated particles with the detector, and its response,
are implemented using the \geant
toolkit~\cite{Allison:2006ve, *Agostinelli:2002hh} as described in
Ref.~\cite{LHCb-PROC-2011-006}.

\section{Event selection}
\label{sec:Selection}

After reconstruction of the \D-meson candidate from two oppositely charged particles, the same event selection is applied to all $\Bm \to D^{(*)} h^-$ channels. Since the neutral pion or photon from the vector \Dstar decay is not reconstructed, partially reconstructed $\Bm \to \Dstar h^-$ decays and fully reconstructed $\Bm \to \D h^-$ decays contain the same reconstructed particles, and thus appear in the same sample. These decays are distinguished according to the reconstructed invariant mass $m(Dh)$, as described in Sec.~\ref{sec:Fit}.

The reconstructed \D-meson candidate mass is required to be within $\pm 25 \mevcc$ of the known \Dz mass~\cite{PDG2016}, which corresponds to approximately three times the mass resolution. 
The kaon or pion originating from the \Bm decay, subsequently referred to as the companion particle, is required to have \pt in the range 0.5--10\gevc and $p$ in the range 5--100\gevc. 
These requirements ensure that the track is within the kinematic coverage of the RICH detectors, which are used to provide particle identification (PID) information. Details of the PID calibration procedure are given in Sec.~\ref{sec:Fit}. 
A kinematic fit is performed to each decay chain, with vertex constraints applied to both the \Bm and \D decay products, and the \D candidate constrained to its known mass~\cite{Hulsbergen:2005pu}. 
Events are required to have been triggered by either the decay products of the signal candidate, or by particles produced elsewhere in the $pp$ collision. Each \Bm candidate is associated to the primary vertex (PV) to which it has the smallest \chisqip, which is quantified as the difference in the vertex fit $\chi^2$ of a given PV reconstructed with and without the considered particle.
The \Bm meson candidates with invariant masses in the interval 4900--5900\mevcc are retained. This range is wider than that considered in Ref.~\cite{LHCb-PAPER-2016-003}, in order to include the partially reconstructed $\Bm \to ([h_{1}^+h_{2}^-]_{D} \piz)_{\Dstar} h^-$ and $\Bm \to ([h_{1}^+h_{2}^-]_{D} \gamma)_{\Dstar} h^-$ decays, which fall at $m(Dh)$ values below the known \Bm meson mass. 

A pair of boosted decision tree (BDT) classifiers, implementing the gradient boost algorithm~\cite{Roe}, is employed to achieve further background suppression. 
The BDTs are trained using simulated \mbox{$\Bm \to [\Km \pip]_{D}\Km$} decays and a background sample of $\Km\pip\Km$ combinations in data with invariant mass in the range 5900--7200\mevcc; the training was also repeated using partially reconstructed \mbox{$\Bm \to ([\Km \pip]_{D}\piz)_{\Dstar}\Km$} and \mbox{$\Bm \to ([\Km \pip]_{D}\piz)_{\Dstar}\Km$} decays, and the difference in performance found to be negligible. No evidence of overtraining was found in the training of either BDT. For the first BDT, background candidates with a reconstructed \D-meson mass more than $30$\mevcc from the known \Dz mass are used in the training. In the second BDT, background candidates with a reconstructed \D-meson mass within $\pm25$\mevcc of the known \Dz mass are used. 
A loose requirement on the classifier response of the first BDT is applied prior to training the second one. This focuses the second BDT training on a background sample enriched with fully reconstructed \D mesons. Both BDT classifier responses are found to be uncorrelated with the $B$-candidate invariant mass.

The input to both BDTs is a set of features that characterise the signal decay. These features can be divided into two categories:
(1) properties of any particle and (2) properties of composite particles only (the \D and \Bm candidates). Specifically:
\begin{enumerate}
\item{$p$, \pt and \chisqip;}
\item{decay time, flight distance, decay vertex quality, radial distance between the decay vertex and the PV, and the angle between the particle's momentum vector and the line connecting the production and decay vertices.}
\end{enumerate} 
In addition, a feature that estimates the imbalance of \pt around the \Bm candidate momentum vector is also used in both BDTs. It is defined as
\begin{equation}
I_{\pt} = \frac{\pt(\Bm) - \Sigma \pt}{\pt(\Bm) + \Sigma \pt}\,,
\end{equation}
where the sum is taken over tracks inconsistent with originating from the PV which lie within a cone around the \Bm candidate, excluding tracks used to make the signal candidate.
The cone is defined by a circle with a radius of 1.5 units in the plane of pseudorapidity and azimuthal angle (expressed in radians). Including the $I_{\pt}$ feature in the BDT training gives preference to \Bm candidates that are either isolated from the rest of the event, or consistent with a recoil against another $b$ hadron. 

Since no PID information is used in the BDT classifier, the efficiency for $\Bm \to D^{(*)}\Km$ and $\Bm \to D^{(*)} \pim$ decays is similar, with insignificant variations arising from small differences in the decay kinematics. 
The criteria applied to the two BDT responses are optimised by minimising the expected statistical uncertainty on $R^{\CP,\piz}$ and $R^{\CP,\gamma}$, as measured with the method described below. 
The purity of the sample is further improved by requiring that all kaons and pions in the \D decay are positively identified by the RICH. This PID selection used to separate the $D\pi$ and $DK$ samples has an efficiency of about 85\% per final-state particle.

Peaking background contributions from charmless decays that result in the same final state as the signal are suppressed by requiring that the flight distance of the \D candidate from the \Bm decay vertex is larger than two times its uncertainty. After the above selections, multiple candidates exist in 0.1\% of the events in the sample.
When more than one candidate is selected, only the candidate with the best \Bm vertex quality is retained. The overall effect of the multiple-candidate selection is negligible.

% $Id: introduction.tex 51723 2014-04-02 12:20:18Z roldeman $

\section{Fit to data}
\label{sec:Fit}

The values of the \CP observables are determined using a binned extended maximum likelihood fit to the data. Distinguishing between \Bp and \Bm candidates, companion particle hypotheses, and the three \D decay product final states, yields 12 independent samples which are fitted simultaneously. 
The total probability density function (PDF) is built from six signal functions, one for each of the \mbox{$\Bm \to \D \pim$}, \mbox{$\Bm \to \D \Km$}, \mbox{$\Bm \to (\D \piz)_{\Dstar} \pim$}, \mbox{$\Bm \to (\D \piz)_{\Dstar} \Km$}, \mbox{$\Bm \to (\D \gamma)_{\Dstar} \pim$}, and \mbox{$\Bm \to (\D \gamma)_{\Dstar} \Km$} decays. In addition, there are functions which describe the combinatorial background components, background contributions from $B$ decays to charmless final states and background contributions from partially reconstructed decays.  
All functions are identical for \Bp and \Bm decays.

\subsection*{\normalsize\boldmath$\Bm \to \D\pim$}

The $\Bm \to D\pim$ signal component is modelled using an asymmetric double-Gaussian-like function
\begin{equation}
f(m) = f_{\text{core}}\text{ exp}\left(\frac{-(m-\mu)^{2}}{2\sigma_c^{2} + (m-\mu)^{2}\alpha_{L,R}}\right) + (1-f_{\text{core}})\text{ exp}\left(\frac{-(m-\mu)^{2}}{2\sigma_{w}^{2}}\right) \label{eq:Cruijff}
\end{equation}
which has a peak position $\mu$ and core width $\sigma_c$, where $\alpha_{L}(m < \mu)$ and \mbox{$\alpha_{R}(m > \mu)$} parameterise the tails. 
The $\mu$ and $\alpha$ parameters are shared across all samples but the core width parameter varies independently for each \D final state. The additional Gaussian function, with a small fractional contribution, is necessary to model satisfactorily the tails of the peak. 

The $\Bm \to D\pim$ decays misidentified as $\Bm \to D\Km$ are displaced to higher mass in the $D\Km$ subsamples. 
These misidentified candidates are modelled by the sum of two Gaussian functions with a common mean, modified to include tail components as in Eq.~\ref{eq:Cruijff}. 
The mean, widths and $\alpha_{R}$ are left to vary freely, while $\alpha_{L}$ is fixed to the value found in simulation.

\subsection*{\normalsize\boldmath$\Bm \to \D\Km$}

In the $D^{(*)0}\Km$ samples, Eq.~\ref{eq:Cruijff} is used for the $\Bm \to D\Km$ signal function. 
The peak position $\mu$ and the two tail parameters $\alpha_{L}$ and $\alpha_{R}$ are shared with the \mbox{$\Bm \to D \pim$} signal function, as are the wide component parameters $f_{\text{core}}$ and $\sigma_{w}$. 
The core width parameter in each \D mode is related to the corresponding $\Bm \to D \pim$ width by a freely varying ratio common to all \D final states. 

Misidentified $\Bm \to D\Km$ candidates appearing in the $D^{(*)0}\pim$ subsamples are described by a fixed shape obtained from simulation, which is later varied to determine a systematic uncertainty associated with this choice.

\subsection*{\normalsize\boldmath$\Bm \to (\D \piz)_{\Dstar} \pim$}

In partially reconstructed decays involving a vector meson, the $\D h^-$ invariant mass distribution depends upon the spin and mass of the missing particle. 
In the case of $\Bm \to (\D \piz)_{\Dstar} \pim$ decays, the missing neutral pion has spin-parity $0^{-}$. The distribution is parameterised by an upward-open parabola, whose range is defined by the kinematic endpoints of the decay. It is convolved with a Gaussian resolution function, resulting in
\begin{equation}
f(m) = \int_a^b \! \left(\mu - \frac{a+b}{2}\right)^2\left(\frac{1-\xi}{b-a}\mu + \frac{b\xi - a}{b - a}\right)e^{-\frac{(\mu - m)^2}{2\sigma^2}}\text{d}\mu \\
\label{eq:RooHORNSdini}\,.
\end{equation}
The resulting distribution has a characteristic double-peaked shape, visible in Figs.~\ref{fig:fit_kpi}$-$\ref{fig:fit_pipi} as the light grey filled regions appearing to the left of the fully reconstructed $B^{-} \to D^{0} h^{-}$ peaks. The lower and upper endpoints of the parabola are $a$ and $b$, respectively, while the relative height of the lower and upper peaks is determined by the $\xi$ term. When $\xi = 1$, both peaks are of equal height, and deviation of $\xi$ from unity accounts for mass-dependent reconstruction and selection efficiency effects. The values of $a$, $b$ and $\xi$ are taken from fits to simulated events, while the convolution Gaussian width $\sigma$ is allowed to vary freely in the mass fit in each \D mode subsample.

Partially reconstructed $\Bm \to (\D \piz)_{\Dstar} \pim$ decays, where the companion pion is misidentified as a kaon, are parameterised with a semiempirical function, formed from the sum of Gaussian and error functions. The parameters of this function are fixed to the values found in fits to simulated events, and are varied to determine the associated systematic uncertainty.

\subsection*{\normalsize\boldmath$\Bm \to (\D \piz)_{\Dstar} \Km$}

Equation~\ref{eq:RooHORNSdini} is also used to describe partially reconstructed $\Bm \to (\D \piz)_{\Dstar} \Km$ decays, where the width $\sigma$ in each of the $D\Km$ samples is related to the $D\pim$ width by a freely varying ratio $r_{\sigma}$, which is shared across all functions describing partially reconstructed decays. All other shape parameters are shared with the $\Bm \to (\D \piz)_{\Dstar} \pim$ function. 

Partially reconstructed $\Bm \to (\D \piz)_{\Dstar} \Km$ decays, where the companion kaon is misidentified as a pion, are parameterised with a semiempirical function, formed from the sum of Gaussian and error functions. The parameters of this function are fixed to the values found in fits to simulated events, and are varied to determine the associated systematic uncertainty.

\subsection*{\normalsize\boldmath$\Bm \to (\D \gamma)_{\Dstar} \pim$}

Partially reconstructed $\Bm \to (\D \gamma)_{\Dstar} \pim$ decays involve a missing particle of zero mass and spin-parity $1^{-}$. The $\D\pim$ invariant mass distribution is described by a parabola exhibiting a maximum, convolved with a Gaussian resolution function. The functional form of this component is
\begin{equation}
f(m) = \int_a^b \! -(\mu-a)(\mu-b)\left(\frac{1-\xi}{b-a}\mu + \frac{b\xi - a}{b - a}\right)e^{-\frac{(\mu-m)^2}{2\sigma^2}}\text{d}\mu
\label{eq:RooHILLdini}\,.
\end{equation}
This distribution exhibits a broad single peak, as opposed to the double-peaked \mbox{$\Bm \to (\D \piz)_{\Dstar} \pim$} distribution described by Eq.~\ref{eq:RooHORNSdini}. In Figs.~\ref{fig:fit_kpi}$-$\ref{fig:fit_pipi}, this component is visible as the wide hatched regions bounded by solid black curves, which appear below the fully reconstructed $B^{-} \to D^{0} h^{-}$ peaks.

The values of $a$, $b$, $\xi$ and $\sigma$ are fixed using fits to simulated events. The clear difference between the invariant mass distributions of $\Bm \to (\D \gamma)_{\Dstar} \pim$ and $\Bm \to (\D \piz)_{\Dstar} \pim$ decays enables their statistical separation, and hence the determination of \CP observables for each mode independently. 

Partially reconstructed $\Bm \to (\D \gamma)_{\Dstar} \pim$ decays where the companion pion is misidentified as a kaon are treated in an equivalent manner to misidentified $\Bm \to (\D \piz)_{\Dstar} \pim$ decays, as described above.

\subsection*{\normalsize\boldmath$\Bm \to (\D \gamma)_{\Dstar} \Km$}

Equation~\ref{eq:RooHILLdini} is also used to describe partially reconstructed $\Bm \to (\D \gamma)_{\Dstar} \Km$ decays, where the width $\sigma$ in each of the $D\Km$ samples is related to the $D\pim$ width by the ratio $r_{\sigma}$. All other shape parameters are shared with the $\Bm \to (\D \gamma)_{\Dstar} \pim$ function. Partially reconstructed $\Bm \to (\D \piz)_{\Dstar} \Km$ decays where the companion kaon is misidentified as a pion are treated in an equivalent manner to misidentified $\Bm \to (\D \piz)_{\Dstar} \Km$ decays.

\subsection*{\normalsize\textbf{Combinatorial background}}

An exponential function is used to describe the combinatorial background. The exponential function is widely used to describe combinatorial backgrounds to $\Bm$ decays in LHCb, and has been validated for numerous different decay modes. Independent and freely varying exponential parameters and yields are used to model this component in each subsample, with the constraint that the \Bp and \Bm yields are required to be equal. The systematic uncertainty associated with this constraint is negligible.

\subsection*{\normalsize\textbf{Charmless background}}

Charmless $\Bm \to h_{1}^+ h_{2}^- h^-$ decays, where $h_{1}^+$, $h_{2}^-$ and $h^-$ each represent a charged kaon or pion, peak at the \Bm mass and cannot be distinguished effectively from the fully reconstructed $\Bm \to \D h^-$ signals in the invariant mass fit.
A Gaussian function is used to model this component, with a $25 \pm 2\mevcc$ width parameter that is taken from simulation; this is about 50\% wider than the $\Bm \to \D h^-$ signal function, due to the application of a $D$ mass constraint in the calculation of the $B$-candidate invariant mass. This constraint improves the invariant mass resolution for signal decays, but worsens it for charmless background contributions.

Partially reconstructed charmless decays of the type $B \to h_{1}^+ h_{2}^- h^- X$, where $X$ is a charged pion, neutral pion or photon that has not been reconstructed, contribute at low invariant mass. Their contributions are fixed to the fully reconstructed charmless components scaled by relative branching fractions~\cite{PDG2016} and efficiencies determined from simulated samples. A parabola with negative curvature convolved with a Gaussian resolution function is used to model this component, with shape parameter values taken from simulation~\cite{Cowan2017239}.

The charmless contribution is interpolated from fits to the \Bm mass spectrum in both the lower and upper \D-mass sidebands, without the kinematic fit of the decay chain.
The charmless yields are determined independently for \Bp and \Bm candidates and are then fixed in the analysis. Their uncertainties contribute to the systematic uncertainties of the final results. 
The largest charmless contribution is in the $\Bm \to [\pip \pim]_{D}\Km$ mode, which has a yield corresponding to 7\% of the measured signal yield. 

\subsection*{\normalsize\textbf{Partially reconstructed background}}

Several additional partially reconstructed $b$-hadron decays contribute at low invariant mass values.
The dominant contributions are from $\Bm \to \D h^- \piz$ and $\Bzb \to (\D \pip)_{\Dstarp} \pim$ decays, where a neutral pion or positively charged pion is missed in the reconstruction.\footnote{ When considering partially reconstructed background contributions, the assumption is made that the production fractions $f_{u}$ and $f_{d}$ are equal.}
The invariant mass distribution of these sources depends upon the spin and mass of the missing particle, as with the $\Bm \to \Dstar h^-$ signals.
In both cases, the missing particle has spin-parity $0^{-}$, such that the $D h^-$ distribution is parameterised using Eq.~\ref{eq:RooHORNSdini}, with shape parameter values taken from simulation. The Dalitz structure of $\Bm \to \D h^- \piz$ decays is modelled using \textsc{Laura++}~\cite{Laura++}.

Decays in which a particle is missed and a companion pion is misidentified as a kaon are parameterised with a semiempirical function, formed from the sum of Gaussian and error functions. The parameters of each partially reconstructed function are fixed to the values found in fits to simulated events, and are varied to determine the associated systematic uncertainty. The yields of the $\Bm \to \D \pim \piz$ and $\Bm \to \D \Km \piz$ contributions vary independently in each subsample, with a \CP asymmetry that is fixed to zero in the case of the favoured mode but allowed to vary freely in the GLW samples. The yields of the $\Bzb \to (\D \pip)_{\Dstarp} \pim$ and $\Bzb \to (\D \pip)_{\Dstarp} \Km$ contributions, where the $\pip$ is not reconstructed, are fixed relative to the corresponding $\Bm \to \D \pim$ yields using branching fractions~\cite{PDG2016,Aubert:2006cd,Aubert:2005yt} and efficiencies derived from simulation. Their \CP asymmetries are fixed to zero in all subsamples as no \CP violation is expected. 

Further contributions from partially reconstructed \mbox{$\Bm \to (\D \piz/\gamma)_{\Dstar} h^- \piz$} and \mbox{$\Bzb \to (\D \pip)_{\Dstarp} h^- \piz$} decays occur at the lowest values of invariant mass, where two particles are not reconstructed. These decays are described by the sum of several parabolas convolved with resolution functions according to Eqs.~\ref{eq:RooHORNSdini} and~\ref{eq:RooHILLdini}, with shape parameters fixed to the values found in fits to simulated samples. The yields and \CP asymmetries of these contributions vary freely in each subsample.     

Colour-suppressed $\Bz \to \D h^- \pip$ and $\Bz \to \Dstar h^- \pip$ decays also contribute to the background. The rates of these small contributions are fixed relative to their corresponding colour-favoured mode yields using the known relative branching fractions~\cite{PDG2016,Alam:1994bi,LHCb-PAPER-2014-070,Satpathy:2002js,Csorna:2003bw}. In the $\Bm \to [\Kp \Km]_{D} h^-$ samples, $\Lb \to [p^+ \Km \pip]_{\Lc} h^-$ decays contribute to the background when the pion is missed and the proton is misidentified as the second kaon. 
The wide function describing this component is fixed from simulation, but the yield in the $\Bm \to [\Kp \Km]_{D} \pim$ subsample varies freely. 
The $\Lb \to [p^+ \Km \pip]_{\Lc} \Km$ yield is constrained using a measurement of $\mathcal{B}(\Lb\to\Lc\Km)/\mathcal{B}(\Lb\to\Lc\pim)$~\cite{Aaij:2013pka}. In both the $\Bm \to [\Kp \Km]_{D} K^-$ and $\Bm \to [\pip \pim]_{D} K^-$ samples, $\Bs \to \D \Km \pip$ decays in which the companion pion is missed contribute to the background. The function describing this component is fixed from fits to simulated samples generated according the the Dalitz model in Ref.~\cite{LHCb-PAPER-2014-036,Laura++}, and the yield is constrained relative to the corresponding $\Bm \to \D \pim$ mode yield scaled by branching fractions~\cite{PDG2016,Aubert:2006cd,LHCb-PAPER-2013-022}, efficiencies determined from simulation, and the relative production rates of $\Bs$ and $\Bz$ mesons at $\sqrt{s} = 7$\tev~\cite{fsfd}. The increase in relative production rate at 13\tev is small~\cite{LHCb-PAPER-2017-001}, and so the 7\tev value is used to describe all data in the analysis.

\subsection*{PID efficiencies}

In the $D^{(*)}\Km$ subsamples, the $\Bm \to D^{(*)} \pim$ cross-feed is determined by the fit to data. 
The $\Bm \to D^{(*)}\Km$ cross-feed into the $D^{(*)}\pim$ subsamples is not well separated from background,
so the expected yield is determined by a PID calibration procedure using approximately 20 million $\Dstarp \to [\Km \pip]_{D} \pip$ decays. 
The reconstruction of this decay is performed using kinematic variables only, and thus provides a pure sample of \Kmp and \pipm particles unbiased in the PID variables. 
The PID efficiency is parameterised as a function of particle momentum and pseudorapidity, as well as the charged-particle multiplicity in the event. 
The effective PID efficiency of the signal is determined by weighting the calibration sample such that the distributions of these variables match those of selected $\Bm \to \Dz \pim$ signal decays. It is found that 71.2\% of $\Bm \to \D\Km$ decays pass the companion kaon PID requirement, with negligible statistical uncertainty due to the size of the calibration sample; the remaining 28.8\% cross-feed into the $\Bm \to D^{(*)} \pim$ sample. 
With the same PID requirement, approximately 99.5\% of the \mbox{$\Bm \to \D \pim$} decays are correctly identified.
These efficiencies are also taken to represent \mbox{$\Bm \to (\D \piz)_{\Dstar} h^-$} and \mbox{$\Bm \to (\D \gamma)_{\Dstar} h^-$} signal decays in the fit, since the companion kinematics are similar across all decay modes considered.
The related systematic uncertainty is determined by the size of the signal samples used, and thus increases for the lower yield modes. The systematic uncertainty ranges from 0.1\% in $\Bm \to [\Km\pip]_{D}\Km$ to 0.4\% in $\Bm \to [\pip\pim]_{D}\Km$. 

\subsection*{Production and detection asymmetries}

In order to measure \CP asymmetries, the detection asymmetries for \Kpm and \pipm mesons must be taken into account. 
A detection asymmetry of $(-0.87 \pm 0.17)$\% is assigned for each kaon in the final state, primarily due to the fact that the nuclear interaction length of \Km mesons is shorter than that of \Kp mesons. 
It is computed by comparing the charge asymmetries in $\Dm\to\Kp\pim\pim$ and $\Dm\to\KS\pim$ calibration samples, weighted to match the kinematics of the signal kaons. 
The equivalent asymmetry for pions is smaller $(-0.17 \pm 0.10)$\%~\cite{LHCb-PAPER-2016-054}. 
The \CP asymmetry in the favoured $\Bm \to [\Km \pip]_{D}\pim$ decay is fixed to $(+0.09 \pm 0.05)\%$, calculated from current knowledge of \g and $r_B$ in this decay~\cite{LHCb-PAPER-2016-032}, with no assumption made about the strong phase, $\delta_{B}^{D\pi}$.
This enables the effective production asymmetry, $A^{\rm eff}_{\Bpm}$, to be measured and simultaneously subtracted from the charge asymmetry measurements in other modes. 

\subsection*{Yields and selection efficiencies}

\begin{figure}[ht]
  \begin{center}
    \includegraphics*[width=0.99\textwidth]{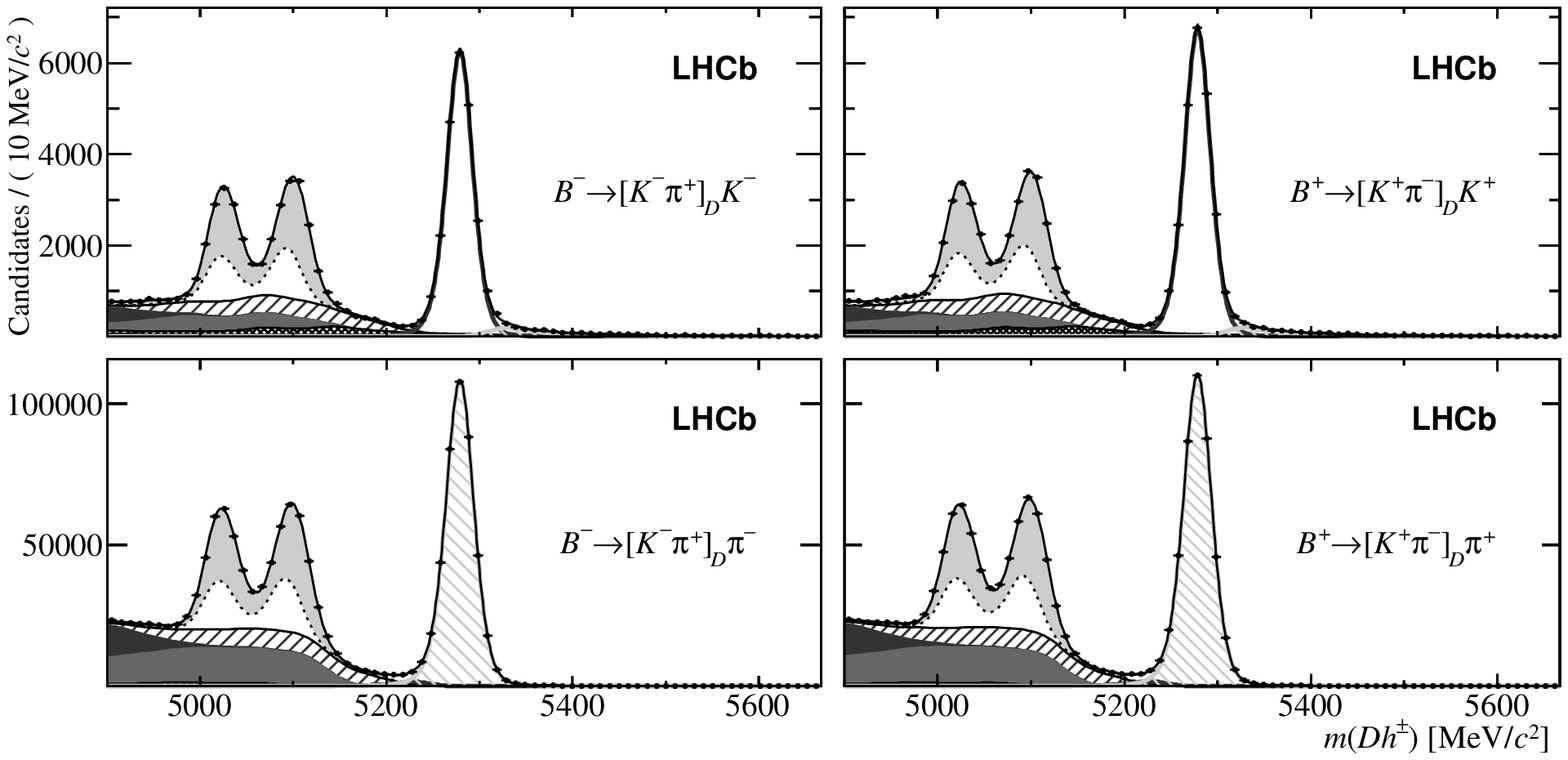}\\
    \hspace*{1cm}\includegraphics[scale=0.35,valign=t]{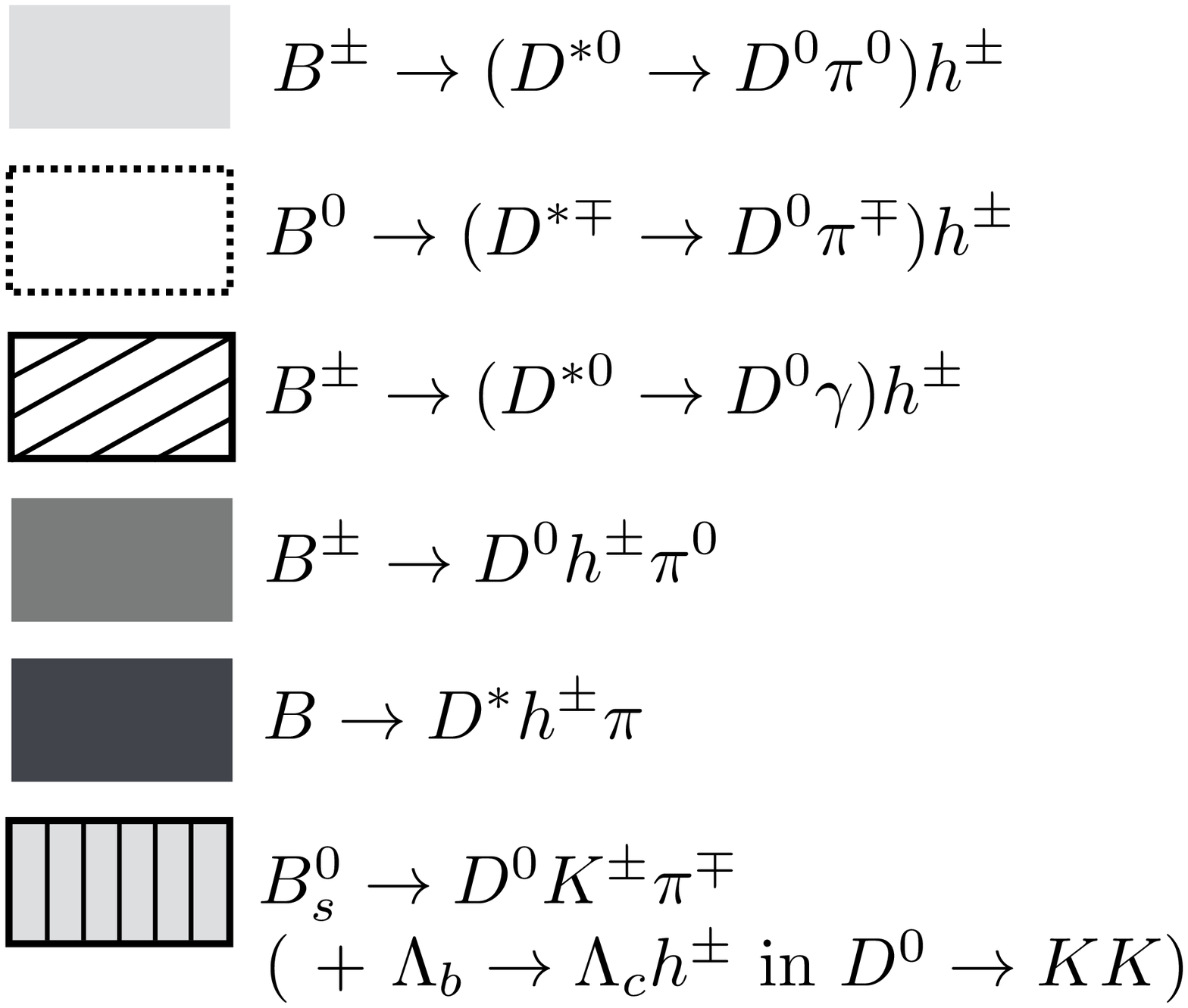}
      \includegraphics[scale=0.35,valign=t]{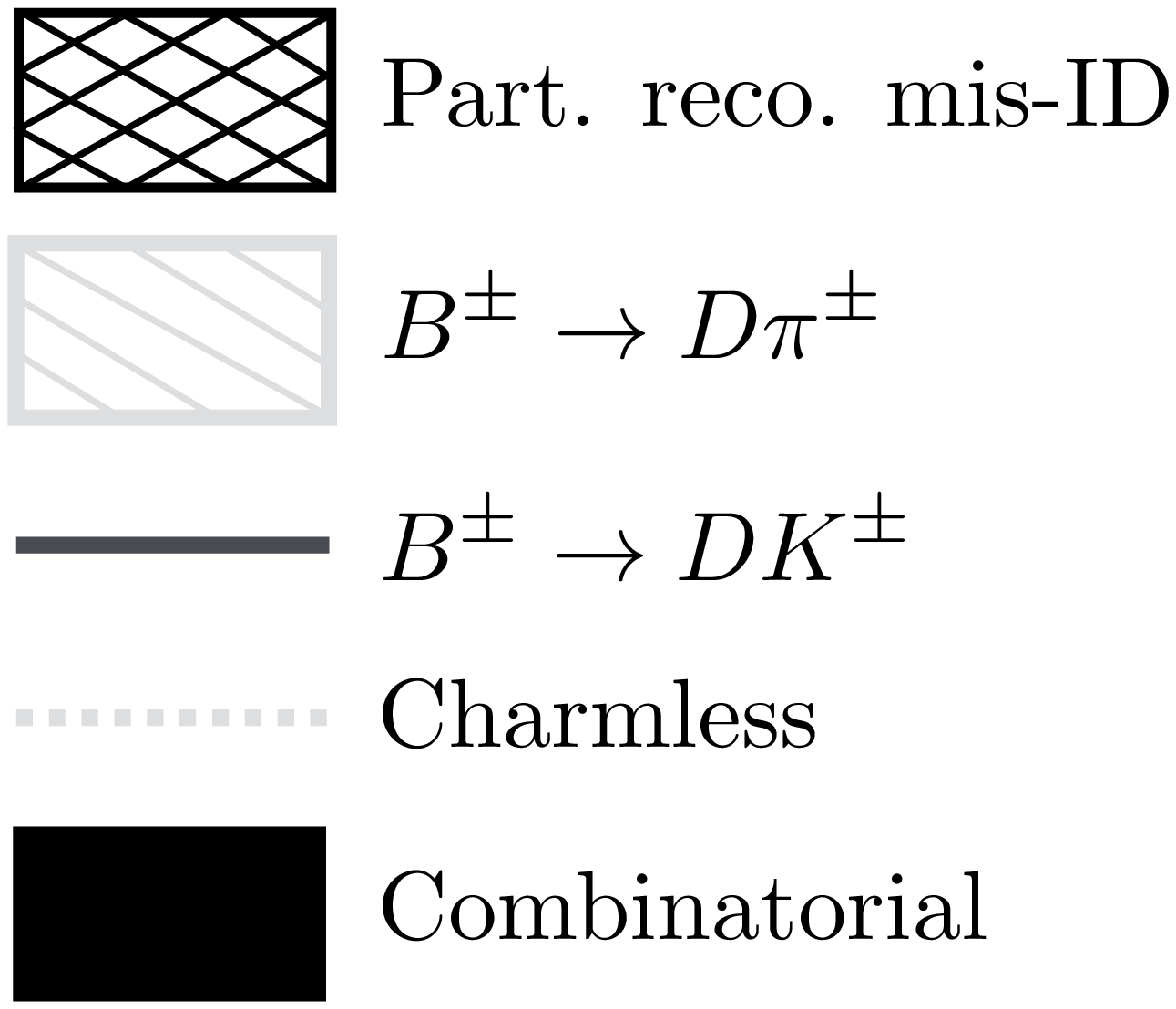}
  \caption{Invariant mass distributions of selected $\Bpm \to [\Kpm \pimp]_{D}h^{\pm}$ candidates, separated
by charge, with $\Bm{\rm(}\Bp{\rm)}$ candidates on the left\,(right). 
The top panels contain the $\Bpm \to D^{(*)0}K^{\pm}$ candidate samples, as defined by a PID requirement on the companion particle.
The remaining candidates are placed in the bottom panels, reconstructed with a pion hypothesis for the companion. 
The result of the fit is shown by the thin solid black line, and each component is listed in the legend. The component referred to as `Part. reco. mis-ID' is the total contribution from all partially reconstructed and misidentified decays. \label{fig:fit_kpi}}
  \end{center}
\end{figure}

\begin{figure}[hb]
  \begin{center}
    \includegraphics*[width=0.99\textwidth]{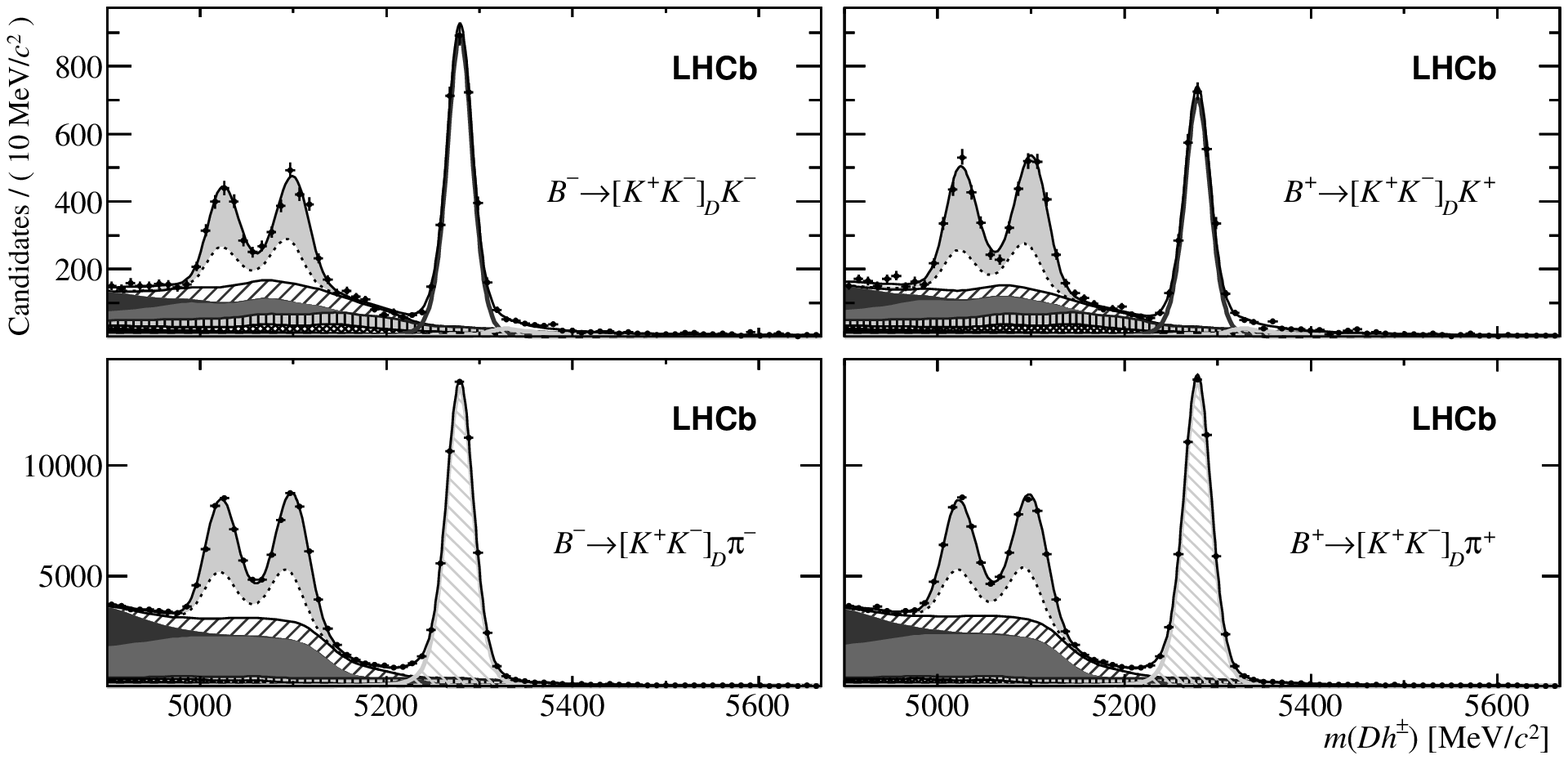}
  \caption{Invariant mass distributions of selected $\Bpm \to [\Kp \Km]_{D}h^{\pm}$ candidates, separated by charge. See Fig.~\ref{fig:fit_kpi} for details of each component.
  \label{fig:fit_kk}}
  \end{center}
\end{figure}

\begin{figure}[ht]
  \begin{center}
    \includegraphics*[width=0.99\textwidth]{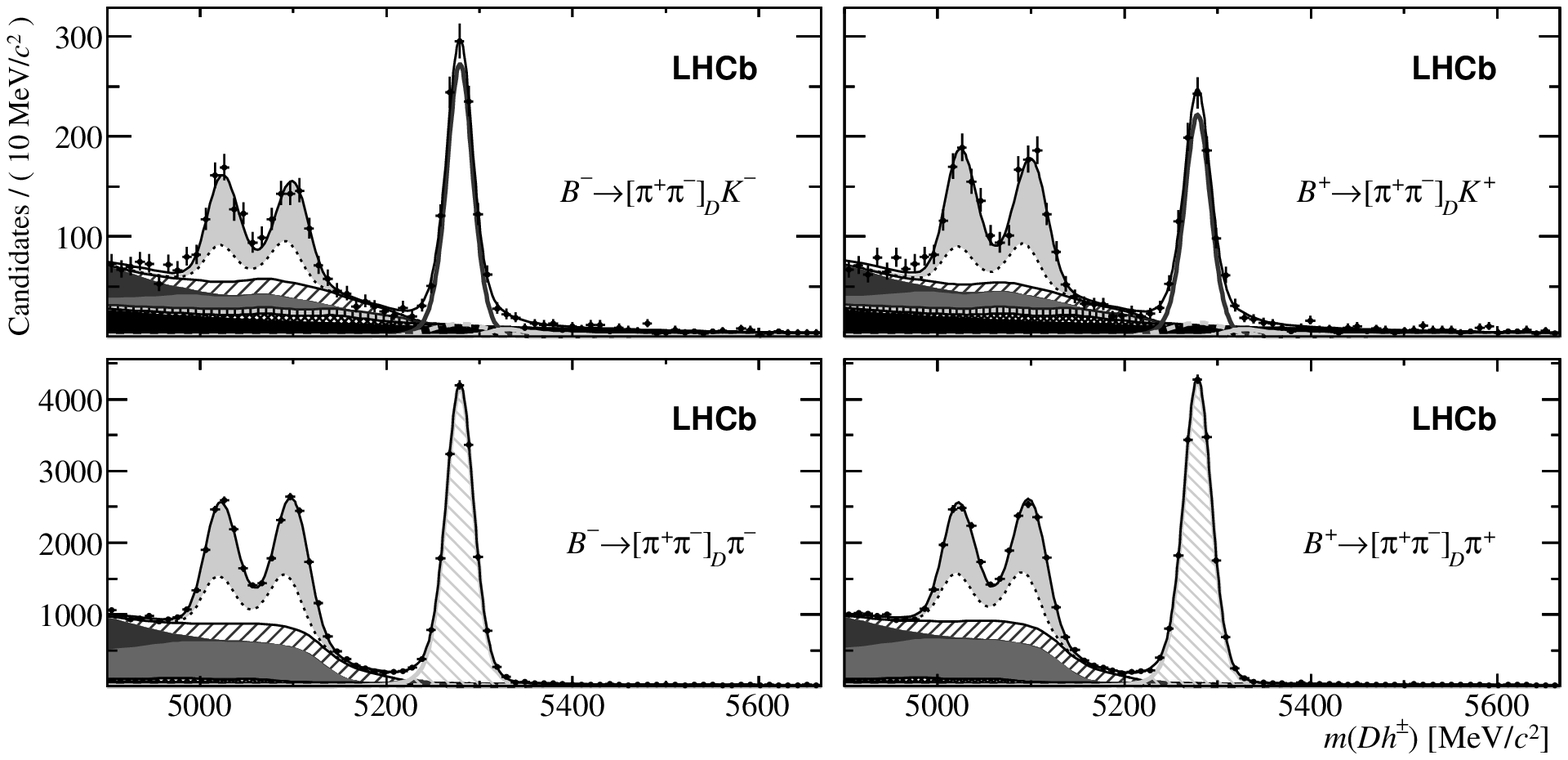}
  \caption{Invariant mass distributions of selected $\Bpm \to [\pip \pim]_{D}h^{\pm}$ candidates, separated by charge. See Fig.~\ref{fig:fit_kpi} for details of each component.
  \label{fig:fit_pipi}}
  \end{center}
\end{figure}

The total yield for each mode is a sum of the number of correctly identified and cross-feed candidates; their values are given in Table~\ref{tab:Yields}. 
The corresponding invariant mass spectra, separated by charge, are shown in \mbox{Figs.~\ref{fig:fit_kpi}$-$\ref{fig:fit_pipi}}.

To obtain the observable $R_{K/\pi}^{K\pi}$ ($R_{K/\pi}^{K\pi,\piz/\gamma}$), which is defined in Table~\ref{tab:AllObservables}, the ratio of yields must be corrected by the relative efficiency with which $\Bm \to \D\Km$ and $\Bm \to \D\pim$ ($\Bm \to \Dstar \Km$ and \mbox{$\Bm \to \Dstar \pim$}) decays are reconstructed and selected. Both ratios are found to be consistent with unity within their assigned uncertainties, which take into account the size of the simulated samples and the imperfect modelling of the relative pion and kaon absorption in the detector material.

To determine the branching fraction \mbox{$\mathcal{B}(\Dstarz \to \Dz \piz)$}, the yields of the \mbox{$\Bm \to (\D \piz)_{\Dstar} \pim$} and \mbox{$\Bm \to (\D \gamma)_{\Dstar} \pim$} modes are corrected for the relative efficiencies of the neutral pion and photon modes as determined from simulation. As both of these modes are partially reconstructed with identical selection requirements, the relative efficiency is found to be unity within its assigned uncertainty, and is varied to determine the associated systematic uncertainty. In the measurement of $\mathcal{B}(\Dstar \to \D \piz)$, the assumption is made that \mbox{$\mathcal{B}(\Dstar \to \D \piz) + \mathcal{B}(\Dstar \to \D \gamma) = 1$}~\cite{PDG2016}.

The branching fraction $\mathcal{B}(\Bm \to \Dstarz \pim)$ is determined from the total $\Bm \to \Dstar \pim$ yield, the total $\Bm \to \D \pim$ yield, the relative efficiencies determined from simulation, and the $\Bm \to \D \pim$ branching fraction~\cite{PDG2016,Aubert:2006cd}. Both the efficiencies and external input branching fraction are varied to determine the associated systematic uncertainty. 

%Signal yields
\begin{table}[!t]
\centering
\begin{tabular}{l r}
\toprule
Mode & Yield \\ \hline
$\Bpm \to [K\pi]_{D} \pipm$ & 862\ 785 $\pm$ \phantom{0}945 \\
$\Bpm \to [KK]_{D} \pipm$ & 105\ 923 $\pm$ \phantom{0}368 \\
$\Bpm \to [\pi\pi]_{D} \pipm$ & \phantom{0}33\ 381 $\pm$ \phantom{0}173 \\
\hline
$\Bpm \to [K\pi]_{D} \Kpm$ & \phantom{0}66\ 987 $\pm$ \phantom{0}326 \\
$\Bpm \to [KK]_{D} \Kpm$ & \phantom{00}8125 $\pm$ \phantom{0}129 \\
$\Bpm \to [\pi\pi]_{D} \Kpm$ & \phantom{00}2571 $\pm$ \phantom{00}70 \\
\hline
$\Bpm \to ([K\pi]_{D} \piz)_{\Dstar} \pipm$ & 519\ 211 $\pm$ 3747 \\
$\Bpm \to ([KK]_{D} \piz)_{\Dstar} \pipm$ & \phantom{0}63\ 742 $\pm$ \phantom{0}460 \\
$\Bpm \to ([\pi\pi]_{D} \piz)_{\Dstar} \pipm$ & \phantom{0}20\ 088 $\pm$ \phantom{0}145 \\
\hline
$\Bpm \to ([K\pi]_{D} \piz)_{\Dstar} \Kpm$ & \phantom{0}40\ 988 $\pm$ \phantom{0}569 \\
$\Bpm \to ([KK]_{D} \piz)_{\Dstar} \Kpm$ & \phantom{00}5725 $\pm$ \phantom{0}165 \\
$\Bpm \to ([\pi\pi]_{D} \piz)_{\Dstar} \Kpm$ & \phantom{00}1804 $\pm$ \phantom{00}52 \\
\hline
$\Bpm \to ([K\pi]_{D} \gamma)_{\Dstar} \pipm$ & 291\ 372 $\pm$ 2103 \\
$\Bpm \to ([KK]_{D} \gamma)_{\Dstar} \pipm$ & \phantom{0}35\ 771 $\pm$ \phantom{0}258 \\
$\Bpm \to ([\pi\pi]_{D} \gamma)_{\Dstar} \pipm$ & \phantom{0}11\ 273 $\pm$ \phantom{00}81 \\
\hline
$\Bpm \to ([K\pi]_{D} \gamma)_{\Dstar} \Kpm$ & \phantom{0}22\ 752 $\pm$ \phantom{0}316 \\
$\Bpm \to ([KK]_{D} \gamma)_{\Dstar} \Kpm$ & \phantom{00}2520 $\pm$ \phantom{0}245 \\
$\Bpm \to ([\pi\pi]_{D} \gamma)_{\Dstar} \Kpm$ & \phantom{000}794 $\pm$ \phantom{00}77 \\
\end{tabular}\captionof{table}{Signal yields as measured in the fit to the data.  \label{tab:Yields}}
\end{table}

\FloatBarrier

\section{Systematic uncertainties}
\label{sec:Systematics}

The 21 observables of interest are free parameters of the fit, and each of them is subject to a set of systematic uncertainties that result from the use of fixed terms in the fit.
The systematic uncertainties associated with using these fixed parameters are assessed by repeating the fit many times, varying the value of each external parameter within its uncertainty according to a Gaussian distribution. The resulting spread (RMS) in the value of each observable is taken as the systematic uncertainty on that observable due to the external source. 
The systematic uncertainties, grouped into six categories, are listed in Tables~\ref{tab:systematics_part_reco} and~\ref{tab:systematics_fully_reco} for the \CP observables measured in a partially reconstructed and fully reconstructed manner, respectively. The systematic uncertainties for the branching fraction measurements are listed in Table~\ref{tab:systematics_BF}. Correlations between the categories are negligible, but correlations within categories are accounted for. The total systematic uncertainties are summed in quadrature. 

The first systematic category, referred to as \textit{PID} in Tables~\ref{tab:systematics_part_reco}$-$\ref{tab:systematics_BF}, accounts for the uncertainty due to the use of fixed PID efficiency values in the fit. The second category \textit{Bkg rate} corresponds to the use of fixed background yields in the fit. For example, the rate of $B^{0} \to D^{*-} \pip$ decays is fixed in the fit using known branching fractions as external inputs. This category also accounts for charmless background contributions, each of which have fixed rates in the fit. The \textit{Bkg func} and \textit{Sig func} categories refer to the use of fixed shape parameters in background and signal functions, respectively; each of these parameters is determined using simulated samples. The category \textit{Sim} accounts for the use of fixed selection efficiencies derived from simulation, for instance the relative efficiency of selecting $\Bm \to (\D \piz)_{\Dstar} \pim$ and $\Bm \to \D \pim$ decays. The final category, \textit{Asym}, refers to the use of fixed asymmetries in the fit. This category accounts for the use of fixed \CP asymmetries and detection asymmetries in the fit, as described earlier.

\begin{table}[h]
\setlength\extrarowheight{5pt}
\renewcommand{\tabcolsep}{5.5pt}
   \begin{center}
   \scriptsize
   \vspace{2mm}
    \caption{Systematic uncertainties for the \CP observables measured in a partially reconstructed manner, quoted as a percentage of the statistical uncertainty on the observable. \label{tab:systematics_part_reco} }

\begin{adjustbox}{max width=\textwidth}
\begin{tabular}{l r r r r r r r r r r r }
\hline
[\%] & $A_{K}^{K\pi,\gamma}$ & $A_{\pi}^{K\pi,\gamma}$ & $A_{K}^{K\pi,\piz}$ & $A_{\pi}^{K\pi,\piz}$ & $A_{K}^{CP,\gamma}$ & $A_{\pi}^{CP,\gamma}$ & $A_{K}^{CP,\piz}$ & $A_{\pi}^{CP,\piz}$ & $R^{CP,\gamma}$ & $R^{CP,\piz}$ & $R_{K/\pi}^{K\pi,\piz/\gamma}$ \\
\hline
\textit{PID}
 & 4.0  & 11.4  & 4.4  & 3.8  & 9.1  & 5.0  & 4.7  & 4.4  & 22.0  & 16.9  & 74.8 \\
\textit{Bkg rate}
 & 3.5  & 1.6  & 3.2  & 3.6  & 40.8  & 3.5  & 16.5  & 5.7  & 114.0  & 41.9  & 180.3 \\
\textit{Bkg func}
 & 8.9  & 1.0  & 3.7  & 0.7  & 24.4  & 1.6  & 27.1  & 1.3  & 42.6  & 25.0  & 417.3 \\
\textit{Sig func}
 & 4.8  & 3.9  & 2.9  & 3.9  & 10.9  & 3.6  & 3.7  & 4.3  & 24.6  & 13.8  & 148.4 \\
\textit{Sim}
 & 3.1  & 1.1  & 2.1  & 1.9  & 6.5  & 0.9  & 4.3  & 2.9  & 23.5  & 15.3  & 153.8 \\
\textit{Asym}
 & 29.9  & 6.8  & 34.1  & 19.4  & 1.0  & 9.4  & 2.2  & 26.1  & 1.4  & 0.6  & 1.9 \\
\hline
Total
 & 32.1  & 14.0  & 35.0  & 20.6  & 50.0  & 11.9  & 32.7  & 27.6  & 128.3  & 55.6  & 507.9 \\
\end{tabular}
\end{adjustbox}
\end{center}
\end{table}

\begin{table}[h]
\setlength\extrarowheight{5pt}
\begin{center}
   %\scriptsize
\vspace{2mm}
    \caption{Systematic uncertainties for the \CP observables measured in a fully reconstructed manner, quoted as a percentage of the statistical uncertainty on the observable. The \textit{Sim} uncertainty on $R_{K/\pi}^{K\pi}$ is due to the limited size of the simulated samples used to determine the relative efficiency for reconstructing and selecting $\Bm \to \D\pim$ and $\Bm \to \D\Km$ decays.  \label{tab:systematics_fully_reco}}
    %\vspace{-3mm}

\begin{tabular}{l r r r r r r r r }
\hline 
[\%]
& $A_{K}^{K\pi}$ & $A_{\pi}^{KK}$ & $A_{K}^{KK}$ & $A_{\pi}^{\pi\pi}$ & $A_{K}^{\pi\pi}$ & $R^{KK}$ & $R^{\pi\pi}$ & $R_{K/\pi}^{K\pi}$ \\
\hline
\textit{PID}
 & 6.0  & 4.3  & 2.0  & 2.7  & 10.3  & 13.8  & 18.8  & 0.0 \\
\textit{Bkg rate}
 & 7.5  & 1.8  & 10.2  & 4.1  & 18.9  & 68.7  & 46.0  & 0.0 \\
\textit{Bkg func}
 & 7.6  & 0.4  & 4.2  & 0.4  & 7.2  & 9.5  & 16.7  & 0.0 \\
\textit{Sig func}
 & 11.1  & 0.9  & 0.8  & 0.9  & 14.3  & 7.9  & 20.9  & 0.0 \\
\textit{Sim}
 & 7.1  & 0.5  & 0.2  & 0.4  & 5.6  & 3.5  & 7.6  & 174.2 \\
\textit{Asym}
 & 37.4  & 52.7  & 3.7  & 31.2  & 2.3  & 0.1  & 0.1  & 0.0 \\
\hline
Total
 & 41.5  & 52.9  & 11.9  & 31.6  & 27.5  & 71.2  & 56.9  & 174.2 \\
\end{tabular}

\end{center}
\end{table}

\begin{table}[h]
\setlength\extrarowheight{5pt}
\begin{center}
   %\scriptsize
\vspace{2mm}
    \caption{Systematic uncertainties for the branching fraction measurements, quoted as a percentage of the statistical uncertainty on the observable.  \label{tab:systematics_BF}}
    \vspace{-3mm}

\begin{tabular}{l r r }
\hline
[\%]
& $\mathcal{B}(\Dstarz \to \Dz \piz)$ & $\mathcal{B}(\Bm \to \Dstarz \pim)$ \\
\hline
\textit{PID}
 & 85.3  & 117.7 \\
\textit{Bkg rate}
 & 364.4  & 672.1 \\
\textit{Bkg func}
 & 52.2  & 29.0 \\
\textit{Sig func}
 & 417.2  & 379.7 \\
\textit{Sim}
 & 295.4  & 509.3 \\
\textit{Asym}
 & 0.2  & 0.3 \\
\hline
Total
 & 635.7  & 932.7 \\
\end{tabular}

\end{center}
\end{table}

% $Id: introduction.tex 51723 2014-04-02 12:20:18Z roldeman $

\section{Results}
\label{sec:results}

The results are
\begin{alignat*}{3}
A_{K}^{K\pi,\gamma} &=\hspace{0.5cm} +0.001 &&\hspace{0.5cm}\pm 0.021  \phantom{1}\text{\stat} &&\hspace{0.5cm}\pm 0.007  \phantom{1}\text{\syst} \\
A_{\pi}^{K\pi,\gamma} &=\hspace{0.5cm} +0.000 &&\hspace{0.5cm}\pm 0.006  \phantom{1}\text{\stat} &&\hspace{0.5cm}\pm 0.001  \phantom{1}\text{\syst} \\
A_{K}^{K\pi,\piz} &=\hspace{0.5cm} +0.006 &&\hspace{0.5cm}\pm 0.012  \phantom{1}\text{\stat} &&\hspace{0.5cm}\pm 0.004  \phantom{1}\text{\syst} \\
A_{\pi}^{K\pi,\piz} &=\hspace{0.5cm} +0.002 &&\hspace{0.5cm}\pm 0.003  \phantom{1}\text{\stat} &&\hspace{0.5cm}\pm 0.001  \phantom{1}\text{\syst} \\
A_{K}^{\CP,\gamma} &=\hspace{0.5cm} +0.276 &&\hspace{0.5cm}\pm 0.094  \phantom{1}\text{\stat} &&\hspace{0.5cm}\pm 0.047  \phantom{1}\text{\syst} \\
A_{\pi}^{\CP,\gamma} &=\hspace{0.5cm} -0.003 &&\hspace{0.5cm}\pm 0.017  \phantom{1}\text{\stat} &&\hspace{0.5cm}\pm 0.002  \phantom{1}\text{\syst} \\
A_{K}^{\CP,\piz} &=\hspace{0.5cm} -0.151 &&\hspace{0.5cm}\pm 0.033  \phantom{1}\text{\stat} &&\hspace{0.5cm}\pm 0.011  \phantom{1}\text{\syst} \\
A_{\pi}^{\CP,\piz} &=\hspace{0.5cm} +0.025 &&\hspace{0.5cm}\pm 0.010  \phantom{1}\text{\stat} &&\hspace{0.5cm}\pm 0.003  \phantom{1}\text{\syst} \\
R^{\CP,\gamma} &=\hspace{0.5cm} \phantom{+}0.902 &&\hspace{0.5cm}\pm 0.087  \phantom{1}\text{\stat} &&\hspace{0.5cm}\pm 0.112  \phantom{1}\text{\syst} \\
R^{\CP,\piz} &=\hspace{0.5cm} \phantom{+}1.138 &&\hspace{0.5cm}\pm 0.029  \phantom{1}\text{\stat} &&\hspace{0.5cm}\pm 0.016  \phantom{1}\text{\syst} \\
R_{K/\pi}^{K\pi,\piz/\gamma} &=\hspace{0.65cm}(7.930 &&\hspace{0.5cm}\pm 0.110  \phantom{1}\text{\stat} &&\hspace{0.5cm}\pm 0.560  \phantom{1}\text{\syst}) \times 10^{-2} \\
\mathcal{B}(\Dstarz \to \Dz \piz) &=\hspace{0.5cm} \phantom{+}0.636 &&\hspace{0.5cm}\pm 0.002  \phantom{1}\text{\stat} &&\hspace{0.5cm}\pm 0.015  \phantom{1}\text{\syst} \\
\mathcal{B}(\Bm \to \Dstarz \pim) &=\hspace{0.65cm}(4.664 &&\hspace{0.5cm}\pm 0.029 \phantom{1}\text{\stat} &&\hspace{0.5cm}\pm 0.268  \phantom{1}\text{\syst}) \times 10^{-3} \\
A_{K}^{K\pi} &=\hspace{0.5cm} -0.019 &&\hspace{0.5cm}\pm 0.005  \phantom{1}\text{\stat} &&\hspace{0.5cm}\pm 0.002  \phantom{1}\text{\syst} \\
A_{\pi}^{KK} &=\hspace{0.5cm} -0.008 &&\hspace{0.5cm}\pm 0.003  \phantom{1}\text{\stat} &&\hspace{0.5cm}\pm 0.002  \phantom{1}\text{\syst} \\
A_{K}^{KK} &=\hspace{0.5cm} +0.126 &&\hspace{0.5cm}\pm 0.014  \phantom{1}\text{\stat} &&\hspace{0.5cm}\pm 0.002  \phantom{1}\text{\syst} \\
A_{\pi}^{\pi\pi} &=\hspace{0.5cm} -0.008 &&\hspace{0.5cm}\pm 0.006  \phantom{1}\text{\stat} &&\hspace{0.5cm}\pm 0.002  \phantom{1}\text{\syst} \\
A_{K}^{\pi\pi} &=\hspace{0.5cm} +0.115 &&\hspace{0.5cm}\pm 0.025  \phantom{1}\text{\stat} &&\hspace{0.5cm}\pm 0.007  \phantom{1}\text{\syst} \\
R^{KK} &=\hspace{0.5cm} \phantom{+}0.988 &&\hspace{0.5cm}\pm 0.015  \phantom{1}\text{\stat} &&\hspace{0.5cm}\pm 0.011  \phantom{1}\text{\syst} \\
R^{\pi\pi} &=\hspace{0.5cm} \phantom{+}0.992 &&\hspace{0.5cm}\pm 0.027  \phantom{1}\text{\stat} &&\hspace{0.5cm}\pm 0.015  \phantom{1}\text{\syst} \\
R_{K/\pi}^{K\pi} &=\hspace{0.65cm}(7.768 &&\hspace{0.5cm}\pm 0.038  \phantom{1}\text{\stat} &&\hspace{0.5cm}\pm 0.066  \phantom{1}\text{\syst}) \times 10^{-2}\,. \\
\end{alignat*}
The results obtained using fully reconstructed $\Bm \to \D h^-$ decays supersede those in Ref.~\cite{LHCb-PAPER-2016-003}, while the $\Bm \to \Dstar h^-$ results are reported for the first time.
The statistical and systematic correlation matrices are given in the appendix. There is a high degree of anticorrelation between partially reconstructed signal and background components in the fit, which all compete for yield in the same invariant mass region. The anticorrelation between the $B^- \to (\D \pi^{0})_{\Dstar} h^-$ and $B^- \to (\D \gamma)_{\Dstar} h^-$ \CP observables is visible in Table~\ref{tab:stat_correlations_part_reco_summary} of the appendix. The presence of such anticorrelations is a natural consequence of the method of partial reconstruction, and limits the precision with which the \CP observables can be measured using this approach.

%A note on A_KK^K compatability in Run 1 vs. Run 2. 
The value of $A_{K}^{KK}$ has increased with respect to the previous result~\cite{LHCb-PAPER-2016-003}, due to a larger value being measured in the $\sqrt{s} = 13$\tev data. The values measured in the independent $\sqrt{s} =$ 7, 8 and 13\tev data sets are consistent within 2.6 standard deviations. All other updated measurements are consistent within one standard deviation with those in Ref.~\cite{LHCb-PAPER-2016-003}.

Observables involving $\D \to \Kp\Km$ and $\D \to \pip\pim$ decays can differ due to \CP violation in the \D decays or acceptance effects. The latest LHCb results~\cite{LHCb-PAPER-2015-055} show that charm \CP-violation effects are negligible for the determination of \g, and that there is also no significant difference in the acceptance for the two modes. Therefore, while separate results are presented for the $\Bm \to \D h^-$ modes to allow comparison with previous measurements, the combined result is most relevant for the determination of \g. The $R^{KK}$ and $R^{\pi\pi}$ observables have statistical and systematic correlations of +0.07 and +0.18, respectively. Taking these correlations into account, a combined weighted average $R^{\CP}$ is obtained
\begin{align*}
R^{\CP} = 0.989 \pm 0.013\stat \pm 0.010\syst\,.
\end{align*}
The same procedure is carried out for the $A_{K}^{KK}$ and $A_{K}^{\pi\pi}$ observables, which have statistical and systematic correlations of +0.01 and +0.05, respectively. The combined average is
\begin{align*}
A_{K}^{\CP} = +0.124 \pm 0.012\stat \pm 0.002\syst\,.
\end{align*}

The observables $R^{\CP,\piz}$ and $A^{\CP,\piz}$ ($R^{\CP,\gamma}$ and $A^{\CP,\gamma}$), measured using partially reconstructed $\Bm \to \Dstar h^-$ decays, can be directly compared with the world average values for $R_{\CP+} \equiv R^{\CP,\piz}$ and $A_{\CP+} \equiv A^{\CP,\piz}$ ($R_{\CP-} \equiv R^{\CP,\gamma}$ and $A_{\CP-} \equiv A^{\CP,\gamma}$) reported by the Heavy Flavor Averaging Group~\cite{HFAG}; agreement is found at the level of 1.5 and 0.4 (1.1 and 1.4) standard deviations, respectively. The values of $R^{\CP,\piz}$ and $A^{\CP,\piz}$ considerably improve upon the world average precision of $R_{\CP+}$ and $A_{\CP+}$, while the measurements of $R^{\CP,\gamma}$ and $A^{\CP,\gamma}$ have a precision comparable to the previous world average.

The value of $R_{K/\pi}^{K\pi,\piz/\gamma}$ is in agreement with, and substantially more precise than, the current world average~\cite{PDG2016,Aubert:2006cd,Aubert:2004hu}. The branching fraction measurements of $\mathcal{B}(\Dstarz \to \Dz \piz)$ and $\mathcal{B}(\Bm \to \Dstarz \pim)$ are found to agree with the current world average values within 0.6  and 1.3 standard deviations, respectively~\cite{PDG2016,Ablikim:2014mww,Aubert:2006cd}. A value for the ratio of branching fractions $\frac{\mathcal{B}(\Bm \to \Dstarz \Km)}{\mathcal{B}(\Bm \to \Dz \Km)}$ is also obtained using the measured results
\begin{equation}
\frac{\mathcal{B}(\Bm \to \Dstarz \Km)}{\mathcal{B}(\Bm \to \Dz \Km)} = \frac{R_{K/\pi}^{K\pi,\piz/\gamma}}{R_{K/\pi}^{K\pi}} \times \frac{\mathcal{B}(\Bm \to \Dstarz \pim)}{\mathcal{B}(\Bm \to \Dz \pim)} = 0.992 \pm 0.077\,,
\end{equation}
where the uncertainty quoted is dominated by systematic uncertainties, and the statistical and systematic correlations between the input observables are fully taken into account. This value is in agreement with, and improves upon, the current world average. The ratios $R_{K/\pi}^{K\pi}$ and $R_{K/\pi}^{K\pi,\piz/\gamma}$ can be interpreted as \mbox{$\Bm \to D^{(*)0}\Km/\Bm \to D^{(*)0}\pim$} branching fraction ratios, in the limit that the suppressed contributions are neglected, which is the same assumption that is made when reporting the results for $\mathcal{B}(\Bm \to \Dstarz\pim)$ and $\frac{\mathcal{B}(\Bm \to \Dstarz \Km)}{\mathcal{B}(\Bm \to \Dz \Km)}$. The branching fraction measurements demonstrate that the method of partial reconstruction is able to measure the $\Bm \to (\D \piz)_{\Dstar} h^-$ and $\Bm \to (\D \gamma)_{\Dstar} h^-$ signals, despite the correlations present in the mass fit.

\section{Conclusion}
\label{sec:conclusions}

World-best measurements of \CP observables in $\Bm \to \D h^-$ decays are obtained with the \D meson reconstructed in the $\Km\pip$, $\Kp\Km$ and $\pip\pim$ final states; these supersede earlier work on the GLW modes presented in Ref.~\cite{LHCb-PAPER-2016-003}. 
Studies of partially reconstructed $\Bm \to \Dstar h^-$ decays are also reported for the first time, where the measurements of \CP observables in $\Bm \to (\D \gamma)_{\Dstar} \Km$ decays are comparable in precision to the current world averages; the equivalent observables measured in $\Bm \to (\D \piz)_{\Dstar} \Km$ decays substantially improve upon the world averages. Evidence of \CP violation in \mbox{$\Bm \to (\D \piz)_{\Dstar} \Km$} decays is found with a statistical significance of 4.3 standard deviations, while the significance of a nonzero value of $A_{K}^{\CP,\gamma}$ is 2.4 standard deviations. The $\Km\pip$ final state, which offers higher sensitivity to $\gamma$ due to larger interference effects~\cite{Atwood:1996ci}, has not been considered in this work, due to the presence of a large background contribution from the poorly understood $\Bsb \to D^{*0} \Kp\pim$ decay.

\begin{figure}[!t]
  \begin{center}
    \includegraphics*[width=0.45\textwidth]{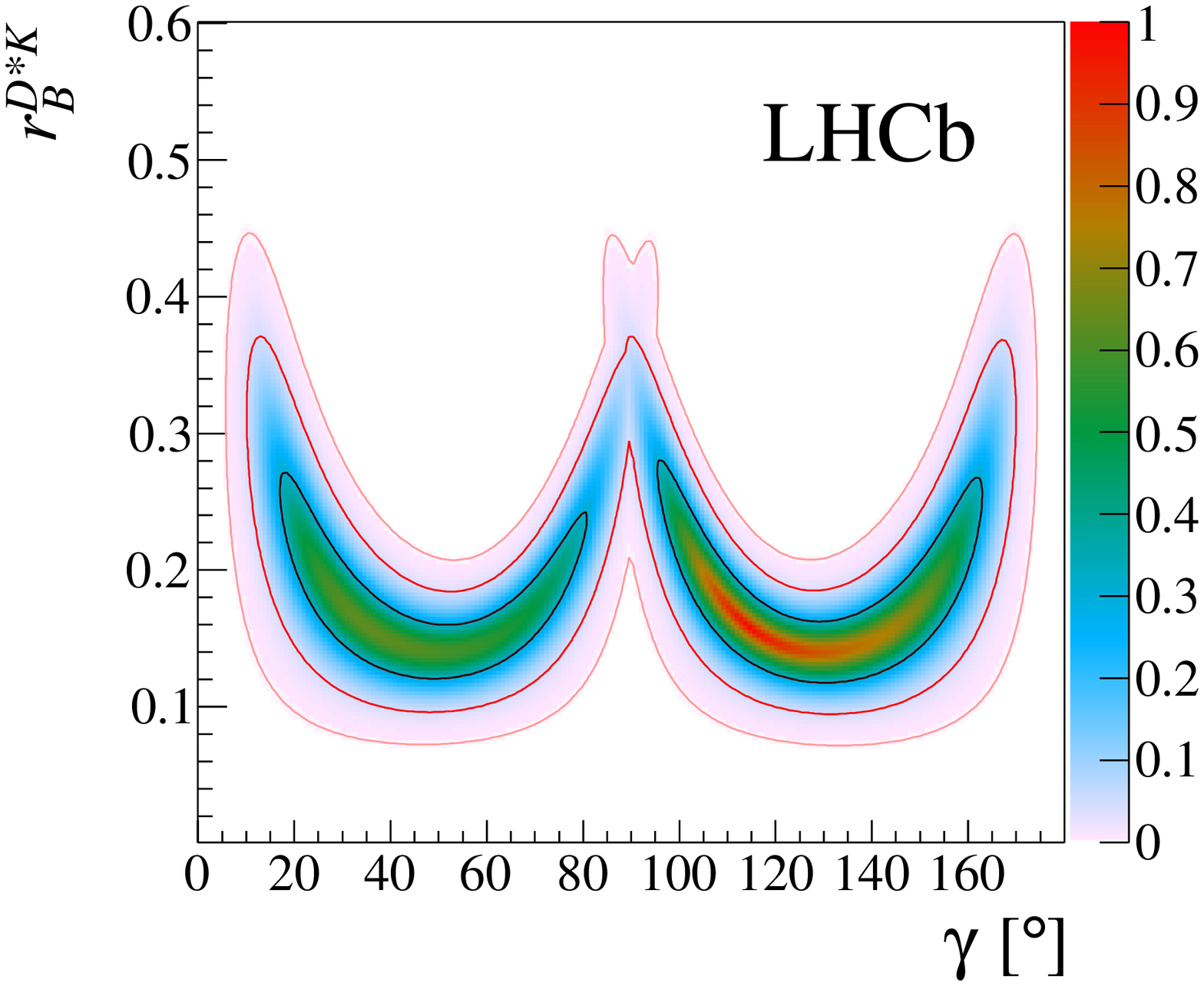}
     \includegraphics*[width=0.45\textwidth]{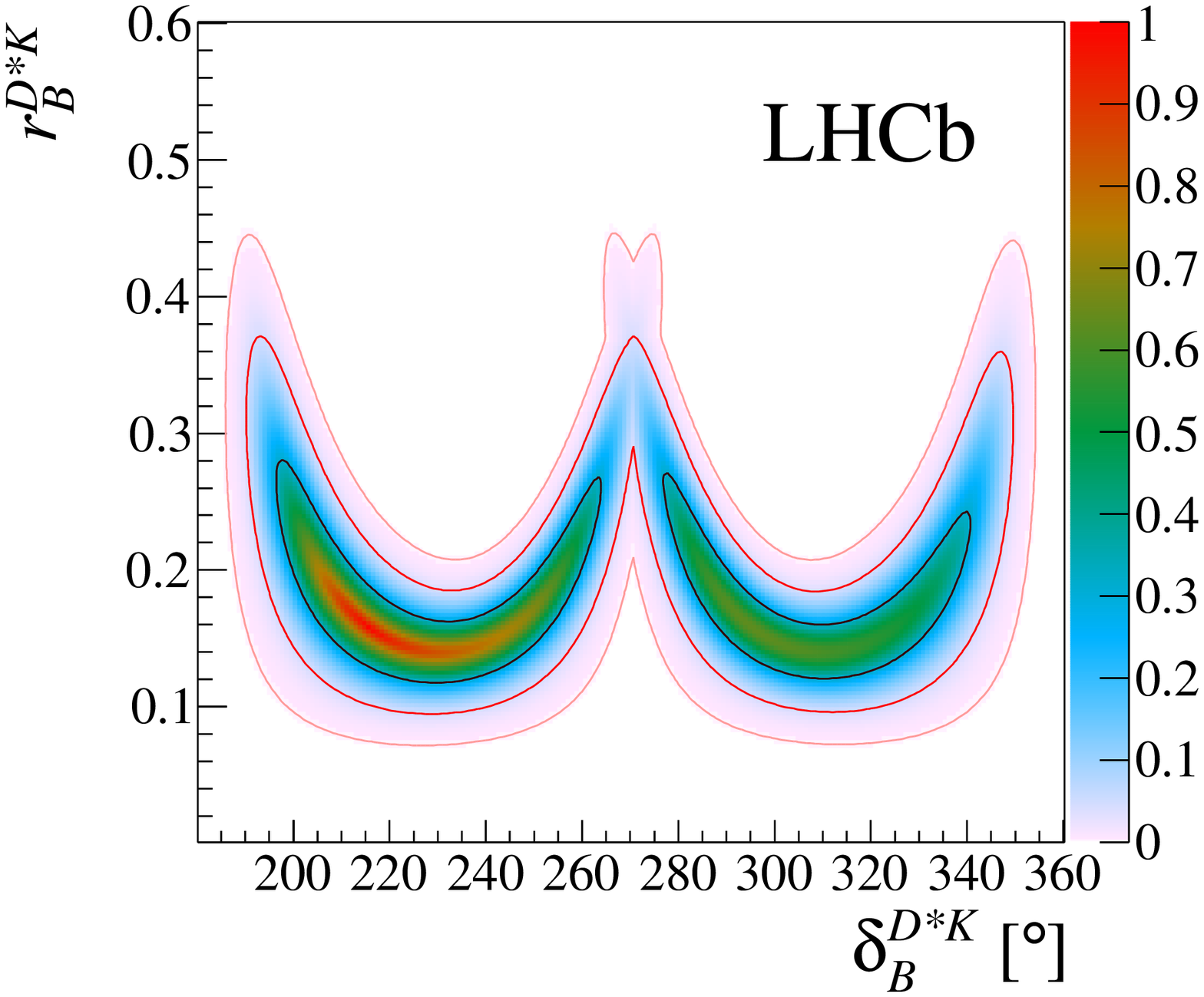}
     \includegraphics*[width=0.45\textwidth]{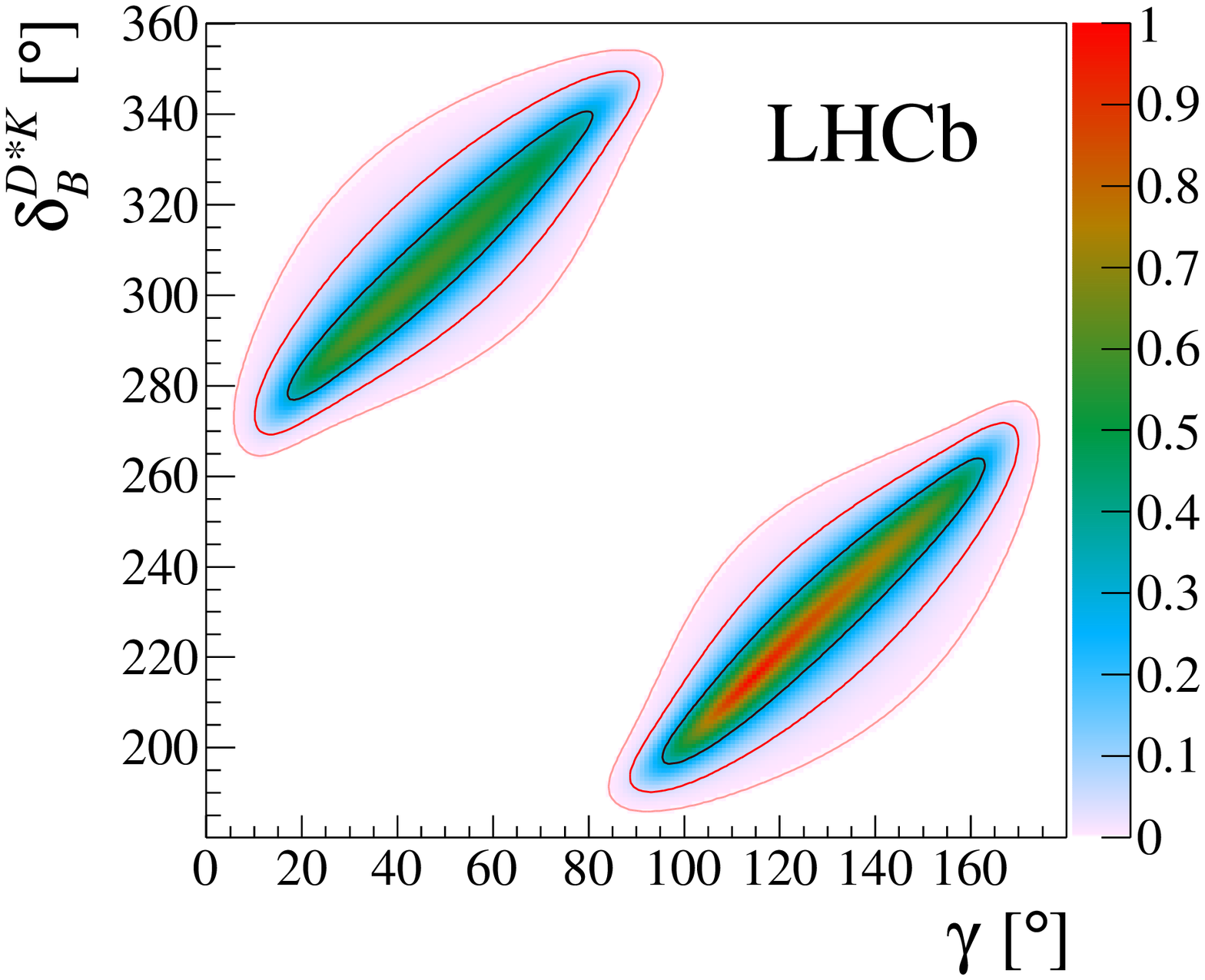}
      \caption{$1\sigma$, $2\sigma$ and $3\sigma$ profile likelihood contours for $r_{B}^{D^{*}\!K}$, $\delta_{B}^{D^{*}\!K}$ and $\gamma$, corresponding to 68.3\%, 95.5\% and 99.7\% confidence level (CL), respectively. The contours are measured using $\Bm \to (\D \piz)_{\Dstar} \Km$ and $\Bm \to (\D \gamma)_{\Dstar} \Km$ decays. The colour scale represents $1-\text{CL}$.  \label{fig:gammadini}}
  \end{center}
  \vspace{-10mm}
\end{figure}

Using the observables $A_{K}^{K\pi,\gamma}$, $A_{K}^{K\pi,\piz}$, $A_{K}^{\CP,\gamma}$, $A_{K}^{\CP,\piz}$, $R^{\CP,\gamma}$ and $R^{\CP,\piz}$ as input, a derivation of the fundamental parameters $r_{B}^{D^{*}\! K}$, $\delta_{B}^{D^{*}\! K}$ and $\gamma$ has been performed using the approach detailed in Ref.~\cite{LHCb-PAPER-2016-032}. The profile likelihood contours at $1\sigma$, $2\sigma$ and $3\sigma$ are shown in Fig.~\ref{fig:gammadini}. The preferred values of $r_{B}^{D^{*}\!K}$ are lower than the current world average values, owing to the fact that the values of $R^{\CP,\gamma}$ and $R^{\CP,\piz}$ measured in this work are below and above unity, respectively, in contrast to the world averages which are both larger than unity~\cite{HFAG}. The preferred values of $\gamma$ and $\delta_{B}^{D^{*}\! K}$ are consistent within 1 standard deviation with the LHCb combination~\cite{LHCb-PAPER-2016-032} and the world average. 

%
% These acknowledgemnets valid from 10-Jul-2017
%
\section*{Acknowledgements}

\noindent We express our gratitude to our colleagues in the CERN
accelerator departments for the excellent performance of the LHC. We
thank the technical and administrative staff at the LHCb
institutes. We acknowledge support from CERN and from the national
agencies: CAPES, CNPq, FAPERJ and FINEP (Brazil); MOST and NSFC
(China); CNRS/IN2P3 (France); BMBF, DFG and MPG (Germany); INFN
(Italy); NWO (The Netherlands); MNiSW and NCN (Poland); MEN/IFA
(Romania); MinES and FASO (Russia); MinECo (Spain); SNSF and SER
(Switzerland); NASU (Ukraine); STFC (United Kingdom); NSF (USA).  We
acknowledge the computing resources that are provided by CERN, IN2P3
(France), KIT and DESY (Germany), INFN (Italy), SURF (The
Netherlands), PIC (Spain), GridPP (United Kingdom), RRCKI and Yandex
LLC (Russia), CSCS (Switzerland), IFIN-HH (Romania), CBPF (Brazil),
PL-GRID (Poland) and OSC (USA). We are indebted to the communities
behind the multiple open-source software packages on which we depend.
Individual groups or members have received support from AvH Foundation
(Germany), EPLANET, Marie Sk\l{}odowska-Curie Actions and ERC
(European Union), ANR, Labex P2IO, ENIGMASS and OCEVU, and R\'{e}gion
Auvergne-Rh\^{o}ne-Alpes (France), RFBR and Yandex LLC (Russia), GVA,
XuntaGal and GENCAT (Spain), Herchel Smith Fund, the Royal Society,
the English-Speaking Union and the Leverhulme Trust (United Kingdom).

\newpage

\appendix
\section*{Appendix: Correlation matrices}
\label{sec:app}

The statistical uncertainty correlation matrices are given in Tables~\ref{tab:stat_correlations_part_reco_summary} and~\ref{tab:stat_correlations_fully_reco_summary} for the \CP observables measured using partially reconstructed and fully reconstructed decays, respectively. The correlations between the systematic uncertainties are provided in Tables~\ref{tab:syst_correlations_part_reco_summary} and \ref{tab:syst_correlations_fully_reco_summary}.

%PR statistical correlation matrix
\begin{table}[ht]
\setlength\extrarowheight{7pt}
\caption{Statistical correlation matrix for the \CP observables measured using partially reconstructed decays.
\label{tab:stat_correlations_part_reco_summary}}
\begin{center}
\scriptsize

\begin{tabular}{l | c c c c c c c c c c c }
& $A_{K}^{K\pi,\gamma}$ & $A_{\pi}^{K\pi,\gamma}$ & $A_{K}^{K\pi,\piz}$ & $A_{\pi}^{K\pi,\piz}$ & $A_{K}^{\CP,\gamma}$ & $A_{\pi}^{\CP,\gamma}$ & $A_{K}^{\CP,\piz}$ & $A_{\pi}^{\CP,\piz}$ & $R^{\CP,\gamma}$ & $R^{\CP,\piz}$ & $R_{K/\pi}^{K\pi,\piz/\gamma}$ \\
\hline
$A_{K}^{K\pi,\gamma}$  & \phantom{$-$}1.00 & $-$0.00 & $-$0.61  & \phantom{$-$}0.01  & \phantom{$-$}0.00  & \phantom{$-$}0.00  & \phantom{$-$}0.00  & \phantom{$-$}0.01  & \phantom{$-$}0.00  & \phantom{$-$}0.00 & $-$0.00 \\
$A_{\pi}^{K\pi,\gamma}$ & $-$0.00  & \phantom{$-$}1.00  & \phantom{$-$}0.04 & $-$0.21  & \phantom{$-$}0.00  & \phantom{$-$}0.01  & \phantom{$-$}0.01  & \phantom{$-$}0.02 & $-$0.00 & $-$0.00  & \phantom{$-$}0.01 \\
$A_{K}^{K\pi,\piz}$ & $-$0.61  & \phantom{$-$}0.04  & \phantom{$-$}1.00  & \phantom{$-$}0.08  & \phantom{$-$}0.00  & \phantom{$-$}0.01  & \phantom{$-$}0.01  & \phantom{$-$}0.02 & $-$0.00 & $-$0.00  & \phantom{$-$}0.00 \\
$A_{\pi}^{K\pi,\piz}$  & \phantom{$-$}0.01 & $-$0.21  & \phantom{$-$}0.08  & \phantom{$-$}1.00  & \phantom{$-$}0.01  & \phantom{$-$}0.02  & \phantom{$-$}0.02  & \phantom{$-$}0.06 & $-$0.00 & $-$0.00  & \phantom{$-$}0.01 \\
$A_{K}^{\CP,\gamma}$  & \phantom{$-$}0.00  & \phantom{$-$}0.00  & \phantom{$-$}0.00  & \phantom{$-$}0.01  & \phantom{$-$}1.00 & $-$0.03 & $-$0.21 & $-$0.02 & $-$0.27  & \phantom{$-$}0.08  & \phantom{$-$}0.01 \\
$A_{\pi}^{\CP,\gamma}$  & \phantom{$-$}0.00  & \phantom{$-$}0.01  & \phantom{$-$}0.01  & \phantom{$-$}0.02 & $-$0.03  & \phantom{$-$}1.00  & \phantom{$-$}0.02 & $-$0.03 & $-$0.01 & $-$0.00  & \phantom{$-$}0.00 \\
$A_{K}^{\CP,\piz}$  & \phantom{$-$}0.00  & \phantom{$-$}0.01  & \phantom{$-$}0.01  & \phantom{$-$}0.02 & $-$0.21  & \phantom{$-$}0.02  & \phantom{$-$}1.00  & \phantom{$-$}0.04 & $-$0.07  & \phantom{$-$}0.12  & \phantom{$-$}0.02 \\
$A_{\pi}^{\CP,\piz}$  & \phantom{$-$}0.01  & \phantom{$-$}0.02  & \phantom{$-$}0.02  & \phantom{$-$}0.06 & $-$0.02 & $-$0.03  & \phantom{$-$}0.04  & \phantom{$-$}1.00 & $-$0.01 & $-$0.00  & \phantom{$-$}0.01 \\
$R^{\CP,\gamma}$  & \phantom{$-$}0.00 & $-$0.00 & $-$0.00 & $-$0.00 & $-$0.27 & $-$0.01 & $-$0.07 & $-$0.01  & \phantom{$-$}1.00 & $-$0.26 & $-$0.14 \\
$R^{\CP,\piz}$  & \phantom{$-$}0.00 & $-$0.00 & $-$0.00 & $-$0.00  & \phantom{$-$}0.08 & $-$0.00  & \phantom{$-$}0.12 & $-$0.00 & $-$0.26  & \phantom{$-$}1.00 & $-$0.15 \\
$R_{K/\pi}^{K\pi,\piz/\gamma}$ & $-$0.00  & \phantom{$-$}0.01  & \phantom{$-$}0.00  & \phantom{$-$}0.01  & \phantom{$-$}0.01  & \phantom{$-$}0.00  & \phantom{$-$}0.02  & \phantom{$-$}0.01 & $-$0.14 & $-$0.15  & \phantom{$-$}1.00 \\
\end{tabular}

\end{center}
\end{table}

%FR statistical correlation matrix
\begin{table}[ht]
\setlength\extrarowheight{7pt}
\caption{Statistical correlation matrix for the \CP observables measured using fully reconstructed decays.
\label{tab:stat_correlations_fully_reco_summary}}
\begin{center}
\scriptsize

\begin{tabular}{l | c c c c c c c c }
& $A_{K}^{K\pi}$ & $A_{\pi}^{KK}$ & $A_{K}^{KK}$ & $A_{\pi}^{\pi\pi}$ & $A_{K}^{\pi\pi}$ & $R^{KK}$ & $R^{\pi\pi}$ & $R_{K/\pi}^{K\pi}$ \\
\hline
$A_{K}^{K\pi}$  & \phantom{$-$}1.00  & \phantom{$-$}0.09  & \phantom{$-$}0.02  & \phantom{$-$}0.05  & \phantom{$-$}0.01  & \phantom{$-$}0.00  & \phantom{$-$}0.00  & \phantom{$-$}0.00 \\
$A_{\pi}^{KK}$  & \phantom{$-$}0.09  & \phantom{$-$}1.00 & $-$0.00  & \phantom{$-$}0.06  & \phantom{$-$}0.02 & $-$0.00  & \phantom{$-$}0.00 & $-$0.00 \\
$A_{K}^{KK}$  & \phantom{$-$}0.02 & $-$0.00  & \phantom{$-$}1.00  & \phantom{$-$}0.02  & \phantom{$-$}0.01 & $-$0.02 & $-$0.00 & $-$0.00 \\
$A_{\pi}^{\pi\pi}$  & \phantom{$-$}0.05  & \phantom{$-$}0.06  & \phantom{$-$}0.02  & \phantom{$-$}1.00 & $-$0.03  & \phantom{$-$}0.00 & $-$0.00 & $-$0.00 \\
$A_{K}^{\pi\pi}$  & \phantom{$-$}0.01  & \phantom{$-$}0.02  & \phantom{$-$}0.01 & $-$0.03  & \phantom{$-$}1.00 & $-$0.00 & $-$0.03 & $-$0.00 \\
$R^{KK}$  & \phantom{$-$}0.00 & $-$0.00 & $-$0.02  & \phantom{$-$}0.00 & $-$0.00  & \phantom{$-$}1.00  & \phantom{$-$}0.07 & $-$0.31 \\
$R^{\pi\pi}$  & \phantom{$-$}0.00  & \phantom{$-$}0.00 & $-$0.00 & $-$0.00 & $-$0.03  & \phantom{$-$}0.07  & \phantom{$-$}1.00 & $-$0.17 \\
$R_{K/\pi}^{K\pi}$  & \phantom{$-$}0.00 & $-$0.00 & $-$0.00 & $-$0.00 & $-$0.00 & $-$0.31 & $-$0.17  & \phantom{$-$}1.00 \\
\end{tabular}

\end{center}
\end{table}

%PR systematic correlation matrix
\begin{table}
\setlength\extrarowheight{7pt}
\caption{Systematic uncertainty correlation matrix for the \CP observables measured using partially reconstructed decays.
\label{tab:syst_correlations_part_reco_summary}}
\begin{center}
\scriptsize

\begin{tabular}{l| c c c c c c c c c c c }

& $A_{K}^{K\pi,\gamma}$ & $A_{\pi}^{K\pi,\gamma}$ & $A_{K}^{K\pi,\piz}$ & $A_{\pi}^{K\pi,\piz}$ & $A_{K}^{\CP,\gamma}$ & $A_{\pi}^{\CP,\gamma}$ & $A_{K}^{\CP,\piz}$ & $A_{\pi}^{\CP,\piz}$ & $R^{\CP,\gamma}$ & $R^{\CP,\piz}$ & $R_{K/\pi}^{K\pi,\piz/\gamma}$ \\
\hline
$A_{K}^{K\pi,\gamma}$  & \phantom{$-$}1.00  & $-$0.02  & \phantom{$-$}0.76  & $-$0.01  & \phantom{$-$}0.01  & $-$0.22  & \phantom{$-$}0.16  & $-$0.24  & \phantom{$-$}0.04  & $-$0.12  & \phantom{$-$}0.11 \\
$A_{\pi}^{K\pi,\gamma}$  & $-$0.02  & \phantom{$-$}1.00  & \phantom{$-$}0.03  & \phantom{$-$}0.61  & \phantom{$-$}0.02  & \phantom{$-$}0.06  & \phantom{$-$}0.03  & \phantom{$-$}0.21  & $-$0.01  & \phantom{$-$}0.02  & \phantom{$-$}0.05 \\
$A_{K}^{K\pi,\piz}$  & \phantom{$-$}0.76  & \phantom{$-$}0.03  & \phantom{$-$}1.00  & \phantom{$-$}0.14  & \phantom{$-$}0.01  & $-$0.46  & $-$0.08  & $-$0.55  & $-$0.01  & \phantom{$-$}0.03  & $-$0.08 \\
$A_{\pi}^{K\pi,\piz}$  & $-$0.01  & \phantom{$-$}0.61  & \phantom{$-$}0.14  & \phantom{$-$}1.00  & $-$0.02  & \phantom{$-$}0.25  & $-$0.00  & \phantom{$-$}0.31  & $-$0.00  & \phantom{$-$}0.00  & \phantom{$-$}0.05 \\
$A_{K}^{\CP,\gamma}$  & \phantom{$-$}0.01  & \phantom{$-$}0.02  & \phantom{$-$}0.01  & $-$0.02  & \phantom{$-$}1.00  & $-$0.04  & \phantom{$-$}0.24  & $-$0.02  & $-$0.90  & \phantom{$-$}0.47  & $-$0.02 \\
$A_{\pi}^{\CP,\gamma}$  & $-$0.22  & \phantom{$-$}0.06  & $-$0.46  & \phantom{$-$}0.25  & $-$0.04  & \phantom{$-$}1.00  & $-$0.12  & \phantom{$-$}0.82  & $-$0.02  & $-$0.03  & $-$0.10 \\
$A_{K}^{\CP,\piz}$  & \phantom{$-$}0.16  & \phantom{$-$}0.03  & $-$0.08  & $-$0.00  & \phantom{$-$}0.24  & $-$0.12  & \phantom{$-$}1.00  & $-$0.07  & $-$0.14  & $-$0.15  & \phantom{$-$}0.73 \\
$A_{\pi}^{\CP,\piz}$  & $-$0.24  & \phantom{$-$}0.21  & $-$0.55  & \phantom{$-$}0.31  & $-$0.02  & \phantom{$-$}0.82  & $-$0.07  & \phantom{$-$}1.00  & $-$0.00  & \phantom{$-$}0.04  & $-$0.01 \\
$R^{\CP,\gamma}$  & \phantom{$-$}0.04  & $-$0.01  & $-$0.01  & $-$0.00  & $-$0.90  & $-$0.02  & $-$0.14  & $-$0.00  & \phantom{$-$}1.00  & $-$0.62  & $-$0.06 \\
$R^{\CP,\piz}$  & $-$0.12  & \phantom{$-$}0.02  & \phantom{$-$}0.03  & \phantom{$-$}0.00  & \phantom{$-$}0.47  & $-$0.03  & $-$0.15  & \phantom{$-$}0.04  & $-$0.62  & \phantom{$-$}1.00  & \phantom{$-$}0.00 \\
$R_{K/\pi}^{K\pi,\piz/\gamma}$  & \phantom{$-$}0.11  & \phantom{$-$}0.05  & $-$0.08  & \phantom{$-$}0.05  & $-$0.02  & $-$0.10  & \phantom{$-$}0.73  & $-$0.01  & $-$0.06  & \phantom{$-$}0.00  & \phantom{$-$}1.00 \\
\end{tabular}

\end{center}
\end{table}

%FR systematic correlation matrix
\begin{table}
\setlength\extrarowheight{7pt}
\caption{Systematic uncertainty correlation matrix for the \CP observables measured using fully reconstructed decays.
\label{tab:syst_correlations_fully_reco_summary}}
\begin{center}
\scriptsize

\begin{tabular}{l| c c c c c c c c }

& $A_{K}^{K\pi}$ & $A_{\pi}^{KK}$ & $A_{K}^{KK}$ & $A_{\pi}^{\pi\pi}$ & $A_{K}^{\pi\pi}$ & $R^{KK}$ & $R^{\pi\pi}$ & $R_{K/\pi}^{K\pi}$ \\
\hline
$A_{K}^{K\pi}$  & \phantom{$-$}1.00  & $-$0.75  & \phantom{$-$}0.07  & $-$0.75  & $-$0.01  & $-$0.03  & $-$0.02  & $-$0.15 \\
$A_{\pi}^{KK}$  & $-$0.75  & \phantom{$-$}1.00  & \phantom{$-$}0.08  & \phantom{$-$}0.99  & \phantom{$-$}0.05  & \phantom{$-$}0.01  & \phantom{$-$}0.01  & $-$0.24 \\
$A_{K}^{KK}$  & \phantom{$-$}0.07  & \phantom{$-$}0.08  & \phantom{$-$}1.00  & \phantom{$-$}0.11  & \phantom{$-$}0.05  & $-$0.48  & $-$0.12  & \phantom{$-$}0.20 \\
$A_{\pi}^{\pi\pi}$  & $-$0.75  & \phantom{$-$}0.99  & \phantom{$-$}0.11  & \phantom{$-$}1.00  & \phantom{$-$}0.02  & \phantom{$-$}0.00  & $-$0.01  & $-$0.24 \\
$A_{K}^{\pi\pi}$  & $-$0.01  & \phantom{$-$}0.05  & \phantom{$-$}0.05  & \phantom{$-$}0.02  & \phantom{$-$}1.00  & \phantom{$-$}0.01  & \phantom{$-$}0.21  & \phantom{$-$}0.07 \\
$R^{KK}$  & $-$0.03  & \phantom{$-$}0.01  & $-$0.48  & \phantom{$-$}0.00  & \phantom{$-$}0.01  & \phantom{$-$}1.00  & \phantom{$-$}0.18  & \phantom{$-$}0.01 \\
$R^{\pi\pi}$  & $-$0.02  & \phantom{$-$}0.01  & $-$0.12  & $-$0.01  & \phantom{$-$}0.21  & \phantom{$-$}0.18  & \phantom{$-$}1.00  & \phantom{$-$}0.04 \\
$R_{K/\pi}^{K\pi}$  & $-$0.15  & $-$0.24  & \phantom{$-$}0.20  & $-$0.24  & \phantom{$-$}0.07  & \phantom{$-$}0.01  & \phantom{$-$}0.04  & \phantom{$-$}1.00 \\
\end{tabular}

\end{center}
\end{table}
\clearpage

\newpage
\addcontentsline{toc}{section}{References}
\setboolean{inbibliography}{true}
\bibliographystyle{LHCb}
\bibliography{main,LHCb-PAPER}
 
\newpage
\centerline{\large\bf LHCb collaboration}
\begin{flushleft}
\small
R.~Aaij$^{40}$,
B.~Adeva$^{39}$,
M.~Adinolfi$^{48}$,
Z.~Ajaltouni$^{5}$,
S.~Akar$^{59}$,
J.~Albrecht$^{10}$,
F.~Alessio$^{40}$,
M.~Alexander$^{53}$,
A.~Alfonso~Albero$^{38}$,
S.~Ali$^{43}$
% ...

{\footnotesize \it
% ...
$ ^{5}$Clermont Universit{\'e}, Universit{\'e} Blaise Pascal, CNRS/IN2P3, LPC, Clermont-Ferrand, France\\
% ...
$ ^{10}$Fakult{\"a}t Physik, Technische Universit{\"a}t Dortmund, Dortmund, Germany\\
% ...
$ ^{38}$ICCUB, Universitat de Barcelona, Barcelona, Spain\\
$ ^{39}$Universidad de Santiago de Compostela, Santiago de Compostela, Spain\\
$ ^{40}$European Organization for Nuclear Research (CERN), Geneva, Switzerland\\
% ...
$ ^{43}$Nikhef National Institute for Subatomic Physics, Amsterdam, The Netherlands\\
% ...
$ ^{48}$H.H. Wills Physics Laboratory, University of Bristol, Bristol, United Kingdom\\
% ...
$ ^{53}$School of Physics and Astronomy, University of Glasgow, Glasgow, United Kingdom\\
% ...
$ ^{59}$University of Cincinnati, Cincinnati, OH, United States\\
% ...
}
\end{flushleft}

%\newpage
%\input{supplementary}
\end{document}